\documentclass[useAMS,usenatbib]{mn2e}

\usepackage{graphicx,amssymb,bm}
\usepackage{subfig,xfrac}
\usepackage{float, Abbrev}
\usepackage[fleqn]{amsmath}  
\usepackage{color}
\usepackage[draft]{hyperref}

\usepackage{multirow}

\newcommand{\be}{\begin{equation}}
\newcommand{\ee}{\end{equation}}
\newcommand{\ba}{\begin{eqnarray}}
\newcommand{\ea}{\end{eqnarray}}

\newcommand{\mfive}{\log(M(<5$kpc$)/M_\odot)}
\def\simlt{\lower.5ex\hbox{$\; \buildrel < \over \sim \;$}}

\newcommand{\zl}{z_{\rm L}}
\newcommand{\zs}{z_{\rm S}}
\newcommand{\dl}{D_{\rm L}}
\newcommand{\ds}{D_{\rm S}}
\newcommand{\dls}{D_{\rm LS}}

\newcommand{\fig}{\begin{figure} \begin{center}}
\newcommand{\efig}{\end{center}\end{figure} }
\newcommand{\figs}{\begin{figure*}\begin{minipage}{180mm} \begin{center}}
\newcommand{\efigs}{\end{center}\end{minipage}\end{figure*} }

\title[A 4\% statistical measurement of $H_0$]{A 4\% measurement of $H_0$ using the cumulative distribution of strong-lensing time delays in doubly-imaged quasars}
\author[D. Harvey]{David Harvey$^{1}$\thanks{e-mail: {\tt harvey@lorentz.leidenuniv.nl}}\\
$^{1}$Lorentz Institute, Leiden University, Niels Bohrweg 2, Leiden, NL-2333 CA, The Netherlands}
\begin{document}

\date{Accepted ---. Received ---; in original form \today.}

\pagerange{\pageref{firstpage}--\pageref{lastpage}} \pubyear{2017}

\maketitle

\label{firstpage}

\begin{abstract}
In the advent of large scale surveys, individually modelling strong-gravitational lenses and their counterpart time-delays in order to precisely measure $H_0$ will become computationally expensive, and highly complex.
A complimentary approach is to study the cumulative distribution function (CDF) of time-delays where the global population of lenses is modelled along with $H_0$.
In this paper we use a suite of hydro-dynamical simulations to estimate the CDF of time-delays from doubly-imaged quasars for a realistic distribution of lenses. We find that the CDFs exhibit large amounts of halo-halo variance, regulated by the density profile inner slope and the total mass within $5$kpc. With the objective of fitting to data, we compress the CDFs using Principal Component Analysis and fit a Gaussian Processes Regressor consisting of three physical features: the redshift of the lens, $\zl$; the power law index of the halo, $\alpha$, and the mass within $5$kpc, plus four cosmological features. Assuming a flat Universe, we fit our model to 27 doubly-imaged quasars finding   $H_0=71^{+2}_{-3}$km/s/Mpc, $\zl = 0.36_{-0.09}^{+0.2} $, $\alpha=-1.8_{-0.1}^{+0.1}$, $\mfive=11.1_{-0.1}^{+0.1} $, $\Omega_{\rm M} = 0.3_{-0.04}^{+0.04} $ and $\Omega_{\rm \Lambda}=0.7_{-0.04}^{+0.04}$. We compare our estimates of $\zl$ and $\mfive$ to the data and find that within the sensitivity of the data, they are not systematically biased. We generate mock CDFs and find with that the Vera Rubin Observatory (VRO) could measure $\sigma/H_0$ to $<3\%$, limited by the precision of the model. If we are to exploit fully VRO, we require simulations that sample a larger proportion of the lens population, with a variety of feedback models, exploring all possible systematics.
\end{abstract}


\section{Introduction}

The currently accepted standard model of cosmology states that we live in a homogenous and isotropic Universe that is dominated by an unknown energy density that is causing the observed expanding Universe to accelerate \citep{planckParsFinal}. Despite the community wide acceptance of this model, the specific details are becoming increasingly intriguing. Recent measurements of the expansion rate at the current day, i.e Hubble constant $H_0$, using local estimators \citep{riesshubble,Riess2019,cosmograilH0,holycowBirrer,holicowVelocityDisp,holycowWong} are arguably in tension with measurements made from the early Universe \citep{planckParsFinal}. This tension has become a central issue in cosmology, and much discussed in the literature \citep{troubleH0,solveNeutrinoH0}. If we are to understand this tension it is important that we continue to constrain and study $H_0$ using complimentary probes with orthogonal systematics.

First suggested by \cite{refsdalTimeDelay}, strong gravitational lensing time delays have recently become a viable and competitive tool in this quest to constrain $H_0$ (For a review see \cite{timeDelayCosmography,timeDelayCosmographyCourbin}). When the light from a distant source is heavily distorted and bent, multiple images of the same source can be observed. Since the geodesics of a single source take differing paths with differential lengths, the same variable emission of a quasar is observed at different times in each multiple image. This time delay is caused by two factors, the geometrical difference in path length and the differing potential of the lensing galaxy that the light passes through. Given that the path length is defined by a combination of angular diameter distances to the source and lens, the time delay is sensitive to the Hubble constant with little dependence on the other cosmological parameters \citep{cosmologyFromTimeDelays}.

Current measurement of the Hubble constant from gravitational time delays require not only accurate measurements of the time delay \citep{COSMOGRAIL,cosmograilH0,allTimeDelays} but also detailed information about the lens-source configuration in order to constrain the lensing potential. As such it is useful to gain extra information such as the velocity dispersion of the host lens \citep{holicowVelocityDisp,nextgenTimeDelay,suyuVelocity}. These methods have provided extremely tight constraints ($2.4\%$) on $H_0$ from just a few lenses \citep{holicowVelocityDisp,cosmograilH0,holycowWong,holicowShajib}.

Although a promising method to constrain $H_0$, the study of {\it individual} lenses can be challenging, for example
\begin{itemize}
\item  The modelling of each individual lens is time consuming, and requires extra information in order to understand how it is perturbing the time delay. Much progress has been made with the automation of lens modelling \citep{autoLens,neuralNetLens,shajibLensModeling}, however it is not clear how feasible this will be for the number of lenses expected to be observed by the Vera Rubin Observatory (VRO).
\item The time delays are prone to micro-lensing of individual stars in each galaxy \citep{microlensingTimeDelay}.
\item The time delays are prone to environmental effects, line-of-sight structures and mass sheet degeneracies \citep{environmentLoS,environmentLoSB,lineOfSightTimeDelay,lineOfSightTimeDelayA}.
\end{itemize}

\subsection{Statistical studies of time delays}
Work exploiting statistical measurements of time delays have been limited. The first studies aimed to measure the inner density profile of galaxies, noting that the time delay probability density function (PDF) was sensitive to this, and therefore assuming some known Hubble constant, they could be used to infer the global density profile of galaxies  \citep{timeDelayStatProfile}.
However, it was quickly realised that the power in this technique was to directly measure the Hubble constant itself. Thus \cite{oguriTimeDelayStat} proposed a framework whereby it would be possible to combine many time delays via a `reduced time delay', and assuming knowledge of the lens distribution, constrain the Hubble parameter. In this exploratory paper they numerically and analytically calculated the conditional probability of the Hubble constant given an image configuration and then from these they generated a conditional probability of a time delay.  Using this framework, they consequentially estimated the Hubble constant from $\sim 16$ lenses. However the use of analytical approximations limited their work,  stating that not including the {\it full} distribution of galaxies could bias their estimate. Moreover, they also did not consider environmental effects and the bias on the time delay. A follow-up study by \cite{lensedQuasarsForLSST} that specifically looked at how many lensed quasars and lensed supernova the VRO will observe was carried out. They predicted that the VRO should observe roughly 3000 lensed quasars which will result in a $\sim 2\%$ error in $H_0$, again citing limitations in their method due to not accounting for dark substructures, micro-lensing and other effects. This number was subsequently re-estimated following a set of time delay challenges, whereby  \cite{lensedQuasarsForLSSTLiao} suggested that in-fact the VRO could return as low as 400 useful time delays. Finally \cite{cosmologyFromTimeDelays} looked at doing this with a few well observed objects. They found that with just $\sim100$ well studied lenses that had precise mass measurements, they would be able to garner competitive constraints. 

These detailed pieces of work showed that the VRO indeed has the statistical power to constrain $H_0$ to the target precision of $<2\%$. However, these works were all based on analytical prescriptions of galaxies, often assuming a Singular Isothermal Sphere density profile. As such they do not include substructure in the lens that will cause perturbations, the environmental effect of lenses lying in over dense environments and line-of-sight effects.

In this paper we will extend this body of work, calculating the cumulative distribution function (CDF) of time delays for doubly imaged quasars from a suite of state-of-the-art hydro-dynamical simulations. We, therefore do not assume any analytical form, allowing for a completely free-form CDF. Integrating over all possible image configurations, we will also account for environmental and line-of-sight effects. Finally we develop an innovative framework to fit these CDFs to data provide interesting constraints. We focus this study specifically on double imaged quasars for four main reasons:
\begin{itemize}
\item  They are an efficient probe of $H_0$ since there are approximately 10:1 doubles to quads \citep{lensedQuasarsForLSST,fluxRatiosWDM}.
\item For the same reason, in our small simulation volume we have access to many more doubly imaged quasars as quadruply and can therefore make much stronger statements about the future prospects of this technique.
\item Time delays from doubles are less scattered and are less susceptible to large perturbations in the lens than quadruply imaged quasars \citep{oguriTimeDelayStat}, therefore perturbations caused by unknown feedback in the simulations will have a smaller impact on the predicted CDF (making it easier to predict the CDF of double imaged time delays).
\item Doubles have longer delays, leading to a much better precision on their measurements. Many quads have delays shorter than 20 days, hard to measure with the 3 days cadence of the VRO \citep{lensedQuasarsForLSST}.
\end{itemize}

In section \ref{sec:method}, we introduce the theory, how we generate estimates of the time delay and our validation tests. We then construct the full CDF in \ref{sec:fullPDF} followed by introducing the suite of simulations in \ref{sec:nbody}. In section \ref{sec:results} we show our initial results before constructing our model for the CDF and our constraints on $H_0$ in section \ref{sec:constraints}. Finally we predict the sensitivity of the VRO in section \ref{sec:forecasts} and discuss and conclude in section \ref{sec:conc}.

\section{Methodology}\label{sec:method}
\subsection{The strong-lensing time delay calculation}\label{sec:theory}
Here we present a brief review of the technical details of strong gravitational time delays, however for a full review please see \cite{gravitational_lensing}. We begin by defining a unit-less coordinate system for a position in the lens plane, $\xi$, and source plane $\eta$, where the lens is always positioned the origin of the source plane, $x = \xi / \xi_0$ and $y=\eta/\eta_0$, where $\eta_0 = \xi_0 \dl/\ds$, $\xi_0 = 4\pi\left(\frac{\sigma_v}{c}\right)^2\frac{\dl\dls}{\ds}$, and $\dl$,$\dls$ and $\ds$ are the angular diameter distances between the lens and the observer, the lens and the source and the source and the observer respectively. From this it is possible to show that the time delay due to two photons taking two different geodesics is a combination of the time delay due to a difference in  path length and the fact that the two photons pass through the lens potential at different points inducing a gravitational time delay, i.e.
\be
c\Delta t = \xi{_0}^2(1+\zl)\frac{\ds}{\dl\dls}\left[ \phi(x_1,y) - \phi(x_2,y)\right],
\label{eqn:timeDelay}
\ee
where
\be
\phi(x_i, y) = \frac{(x_i - y)^2}{2} - \Psi(x_i),
\ee
and $\Psi$ is the dimensionless lensing potential defined as the integral of the 3D Newtonian potential $\Phi$ along the line-of-sight, i.e.
\be
\Psi(x) =\frac{1}{\xi_0^2} \frac{\dl\dls}{\ds} \frac{2}{c^2} \int\Phi(\xi, z) {\rm d}z.
\label{eqn:lensingPot}
\ee
\fig
\includegraphics[width=0.49\textwidth]{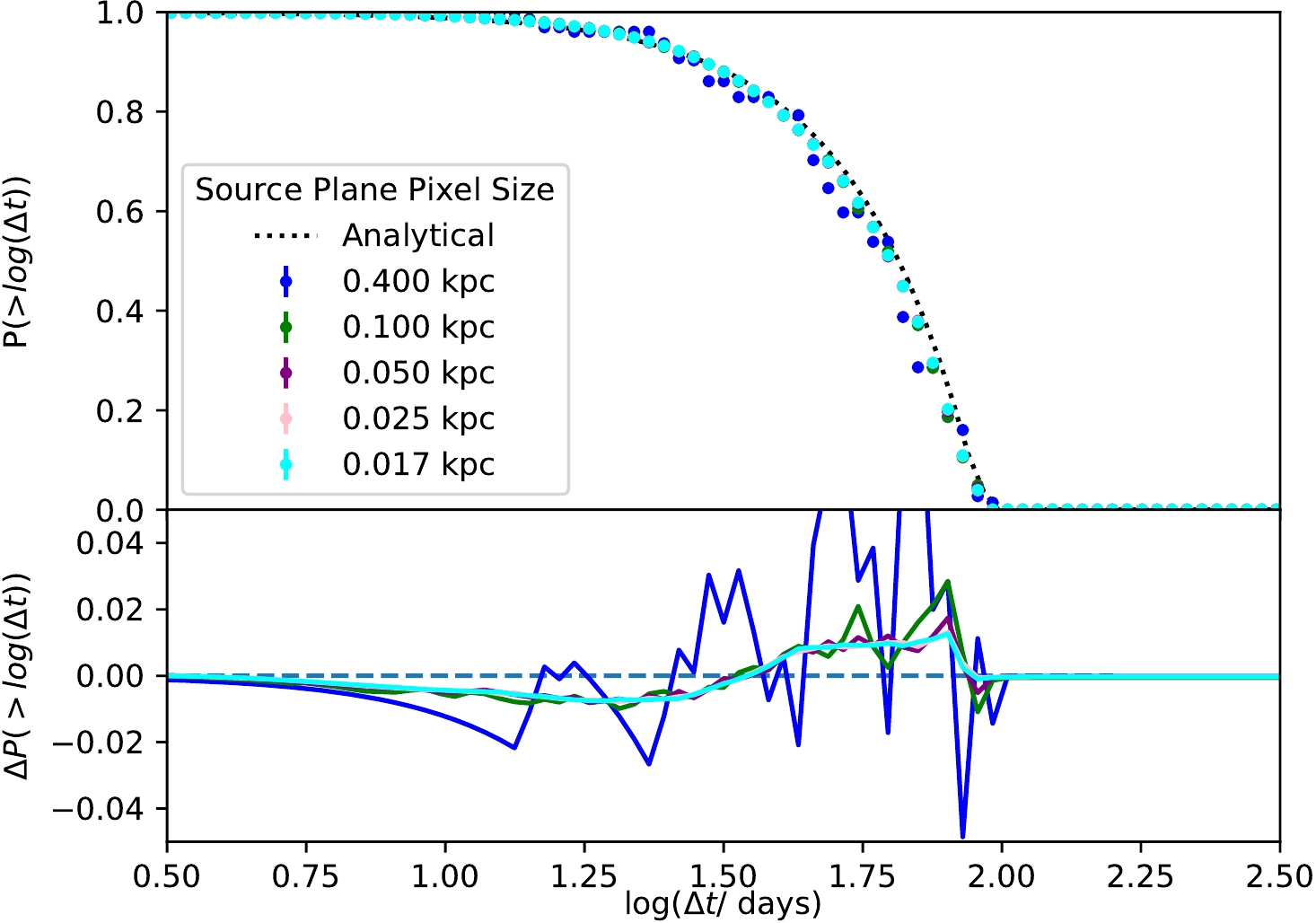}
\includegraphics[width=0.49\textwidth]{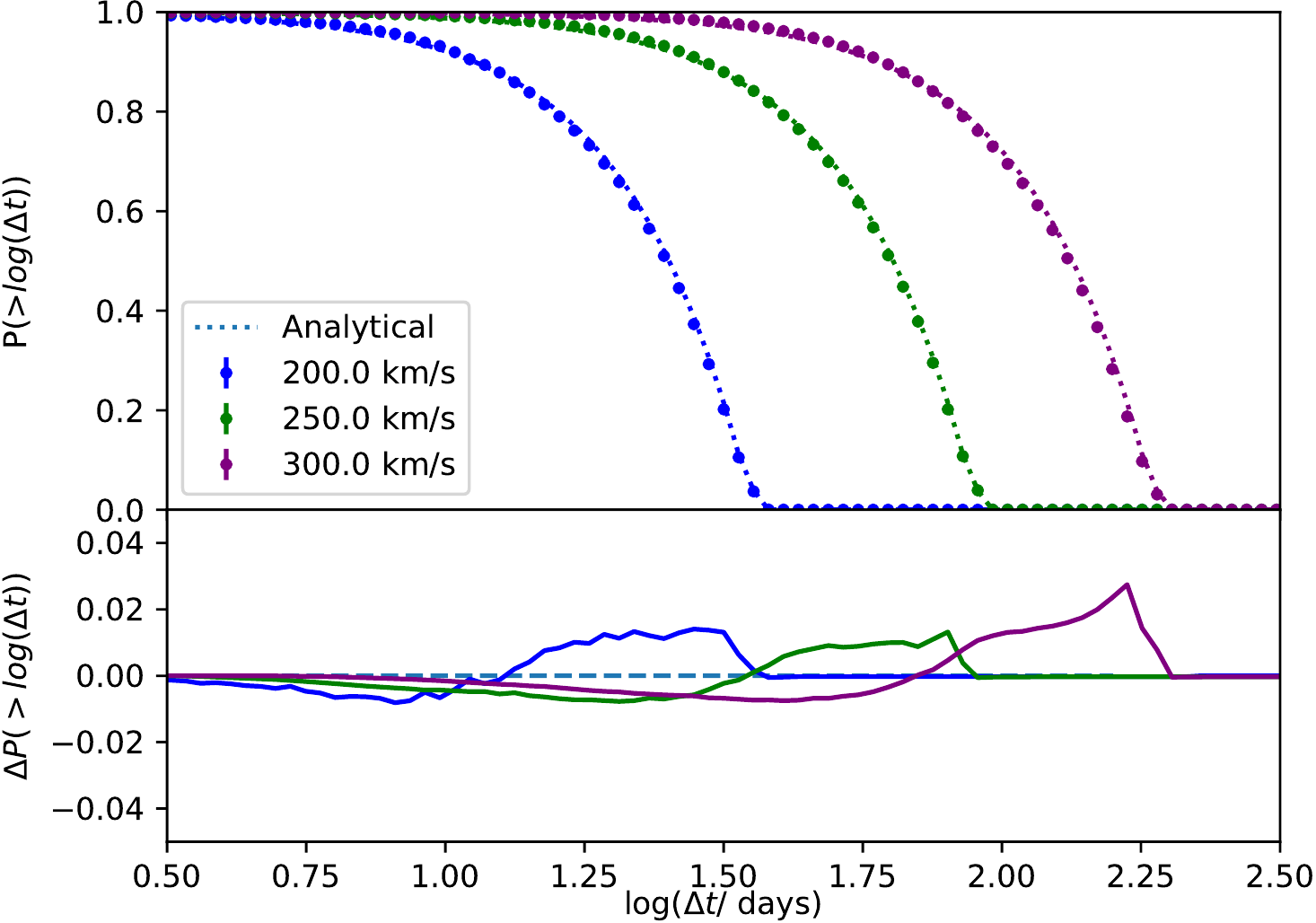}
\caption{\label{fig:SISexample}  Using the known analytical PDF of a Single Isothermal Sphere (SIS) for $\zl=0.2$ and $\zs = 5.$, we carry out two tests on the code. {\it Top, a source plane convergence test: }  We test the sensitivity of the CDF to our pixellised source plane using an SIS with a velocity dispersion of $250$km/s. We test five different source plane resolutions (with a lens plane resolution of $0.1$kpc). The bottom panel shows that CDF converges above a resolution of 0.025kpc. We therefore choose a source plane resolution of $0.025$kpc for the rest of the paper. {\it Bottom: Code accuracy:} We also test the accuracy of the code by analysing three different velocity dispersion of the SIS. The bottom panel shows the difference between the predicted and the analytical value. The adopt the maximal difference of $2\%$ as our theoretical systematic floor.}
\efig
Following equation \eqref{eqn:lensingPot} we can relate the Newtonian potential to the density via the Poisson equation and hence find that the normalised projected surface density, $\kappa$ is the derivative of the lensing potential, i.e. 
\be 
\kappa(x) = \frac{\Sigma(x)}{\Sigma_{\rm crit}} = \frac{1}{2} \bigtriangledown_x \Psi(x),
\ee
where $\Sigma$ is the projected surface density and the critical density is
\be
\Sigma_{\rm crit} = \frac{c^2}{4\pi G}\frac{\ds}{\dl\dls},
\ee
and $G$ is Newton's constant. 
Using this relation between the projected surface density and the time delay, we are able to directly calculate the time delays from a density field. 
 To do this we first calculate $\Psi$ and hence the deflection angle, $\alpha(x)=\bigtriangledown_x \Psi(x)$, for each position in the image plane. Using the lens equation $y = x - \alpha(x)$, we trace a pixellised grid in the image plane to the source plane via the calculated deflection angles and collect all those source plane pixels that have two or more pixels on the image plane, retaining only those source pixels that result in doubly imaged positions.   We follow \cite{lensedQuasarsForLSST} and select only those doublets that have a magnification ratio $\mu_2/\mu_1>0.1$, since in the case where the magnification ratio is  small, a lens with large flux differences would make the second image hard to observe. However, the detection will also depend on the brightness of the quasar and therefore this is an approximation. We then calculate the time arrival surface for the lens and then use equation \eqref{eqn:timeDelay} to calculate the time delay. Before progressing we validate our code on known density profiles.

\subsection{Validation Tests on Single-Isothermal-Spheres}\label{sec:validation}
Following the calculation of the time delays, we validate the code to ensure that it recovers a known distribution of time delays. For a single-isothermal-sphere (SIS), the mass density profile as a function of halo-centric radius, $r$ is
\be
\rho = \frac{\sigma_{\rm v}^2 }{4\pi Gr^2},
\ee
where $\sigma_{\rm v}$ is velocity dispersion. From this it can be shown that the analytical time delay is \citep{timeDelayStatProfile}
\be
c\Delta t = 32 \pi^2 \left(\frac{\sigma}{c}\right)^4 \frac{ \dl\dls}{\ds}(1+\zl)y,
\ee
and is hence just a function of the source position. Integrating over all source positions, $p(\eta, \eta+d\eta)$, the normalised probability is \citep{supernovaTimeDelay}
\be 
p(\eta, \eta +d\eta) = \int_\eta^{\eta+d\eta} 2\frac{\eta}{\eta_0^2}d\eta;~~~~~~~~~\eta < \eta_0,
\ee
and hence
\be
p(\log(\Delta t)) = \frac{2 \ln 10}{\Delta t_{\rm peak}^2} \Delta t^2;~~~~~~~~~\log(\Delta t) < \log(\Delta t_{\rm peak}).
\label{eqn:validationEq}
\ee
Hence the probability $p(\log(\Delta t)) \propto \Delta t^2$.  Indeed \cite{supernovaTimeDelay} suggested that this would hold for any internal density profile slope, $\beta$, such that 
\be
p(\log(\Delta t)) \propto \Delta t^\beta.
\label{eqn:logT}
\ee
We use equation \eqref{eqn:validationEq} to test our code.  We first simulate an SIS halo with a $\sigma_{\rm v} = 250$km/s, with a $\zl=0.2$, a single source plane of $\zs=5$, and calculate the cumulative density function (CDF). In order to estimate the sensitivity of the pixellisation of the source plane to the results we carry out the calculation for different source plane resolutions (assuming a lens plane resolution of $0.1$kpc). The top panel of Figure \ref{fig:SISexample} shows the results of this test. We show the cumulative density function (CDF) (i.e. $p(>\log(\Delta t))$) for five different source plane resolutions and the analytical expectation. We find that the CDF converges at $0.025$kpc where the error in the CDF goes below $\sim2\%$. We therefore choose a source plane resolution of $0.025$kpc for the rest of our analysis. Following this we test the accuracy of our code. We simulate three different velocity dispersions, $\sigma=200, 250, 300$km/s, representing what is expected from massive ellipticals. The bottom panel of Figure \ref{fig:SISexample} shows the results. We find that in all three cases we return the expected distribution to within $2\%$, where this level represents the systematic floor of our analysis.

\section{The full cumulative density function}\label{sec:fullPDF}
Now we have the raw CDF for a single lens-source configuration we construct the full expected CDF. To do this we first consider the observational systematics that will affect this single lens-source plane configuration. This includes line-of-sight structures, mass-sheets and micro-lensing, here we explore them individually.

\fig
\includegraphics[width=0.5\textwidth]{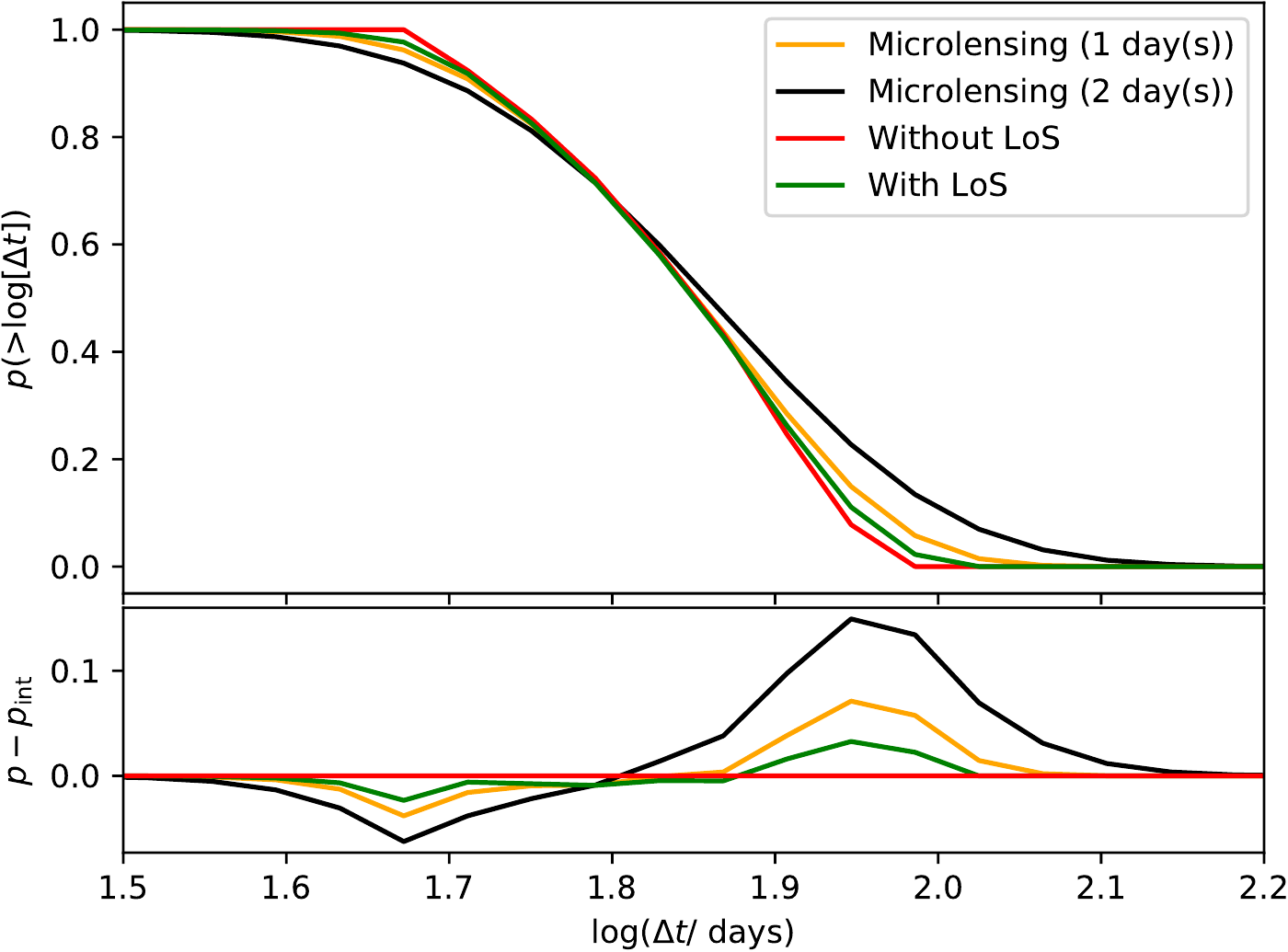}
\caption{\label{fig:LoS}  {\it The impact of line of sight structures \& micro-lensing: }  The observed time delays are perturbed by intervening structures along the line of sight and compact objects within the lens. In order to determine this we convolve the PDF with both numerical and analytical estimates of these effects. Here we show the impact of line-of-sight structures and two levels of micro-lensing, a realistic value of $1$ day and 2 days for a single SIS ($\sigma=250$km/s) lens ($\zl=0.25$) with a source at $\zs=5.0$. The bottom panel shows the difference between each distribution and the intrinsic (red) one. }
\efig

\subsection{Line-of-sight structures \& mass-sheets}\label{sec:los}
It is known the environment and line-of-sight structures perturb geodesics, and hence can shift a single time delay by a non-negligible amount. The impact of a constant mass sheet on the time delay is
\be
\Delta t_{\rm O} = \Delta t_{\rm T} (1-\kappa_{\rm ext}),\label{eqn:losTrue}
\ee
where $\kappa_{\rm ext}$ is the external convergence that perturbs the geodesics, $\Delta t_{\rm O}$ and $\Delta t_{\rm T}$ are the observed time delay that has been perturbed by some external mass and ``true'' time delays, the unperturbed time delay. Hence, $H_0$ will be over-estimated if the external convergence is unaccounted for properly.  Here we explicitly consider two forms of line-of-sight structures: there is the immediate environment of the lens (but not substructure), which on average will be more dense \citep{lensBias}, i.e. mass-sheets; and the uncorrelated structure that the geodesic encounters beyond the lens. We already include the first in the PDFs since we cut out cylinders in the simulation of $1$Mpc (see Section \ref{sec:nbody}), which includes any environment up to a cluster scale (which are not considered in this study), down to the mass resolution of the simulations. The second we consider here. The impact on the PDF on the ``true'', unperturbed time delay, $p_{\rm T}$ is
\be
p(\Delta t_{\rm O}) {\rm d}\Delta t_{\rm O}= p_{\rm T}(\Delta t_{\rm T}) {\rm d}\Delta t_{\rm T} p_\kappa(\kappa_{\rm ext}) {\rm d}\kappa_{\rm ext},
\ee
where $p_\kappa$ is the probability of passing through a halo with a convergence $\kappa$, and $p(\Delta t_{\rm O})$  is the ``observed'' or perturbed PDF.
Substituting in \eqref{eqn:losTrue}, we find
\be
p(\Delta t_{\rm O}) = \int \frac{{\rm d}\Delta t' }{\Delta t'} p_{\rm T}( \Delta t')p_{\kappa}\left( 1- \frac{\Delta t_{\rm O}}{\Delta t'}\right).
\ee
Since we will be carrying out the calculations in $p(\log(\Delta t))$ we convert via the standard Jacobian to get
\be
\begin{split}
p(\log(\Delta t_O)) &=  \ln(10) \int d \log(\Delta t')  \frac{\Delta t_O}{ \Delta t'} \\ &p_T(\log(\Delta t')) p_\kappa(1-\Delta t_O / \Delta t') .
\label{eqn:los}
\end{split}
\ee
 This final Equation \eqref{eqn:los} shows how the distribution of time delays is statistically perturbed by cosmological structures along the line-of-sight in a given cosmology. To incorporate this equation in to our pipeline and estimate $p(\log(\Delta t_O))$  for a given lens-source configuration, we take the estimate of $p(\log(\Delta t_T))$ and for each value we convolve it with the probability of passing through a halo of convergence $p_\kappa$, we then integrate this over the entire distribution of $\log(\Delta t_T)$ to get our final estimate of $\log(\Delta t_O)$. We use the publicly available code, {\sc TurboGL} to calculate the probability density function of the geodesic passing through a halo, $p_\kappa$. {\sc TurboGL} is a numerical code that estimates the cosmological PDF of convergence for any given line-of-sight. It does this by carrying out compound weak lensing to estimate the linear contribution of halos in a $\Lambda$CDM model \citep{turboGL,turboGLa,turboGLb}. We show the results in Figure \ref{fig:LoS} where we show for a simulated SIS lens ($\zl=0.2$, $\zs=5$, $\sigma=250$km/s) the true intrinsic distribution (in red) and the convolved distribution (in green). The bottom panel shows the relative difference between the two. We find that the impact of line-of-sight structures is small even in this extreme case of a source at a redshift of $\zs=5$.

\figs
\includegraphics[width=\textwidth]{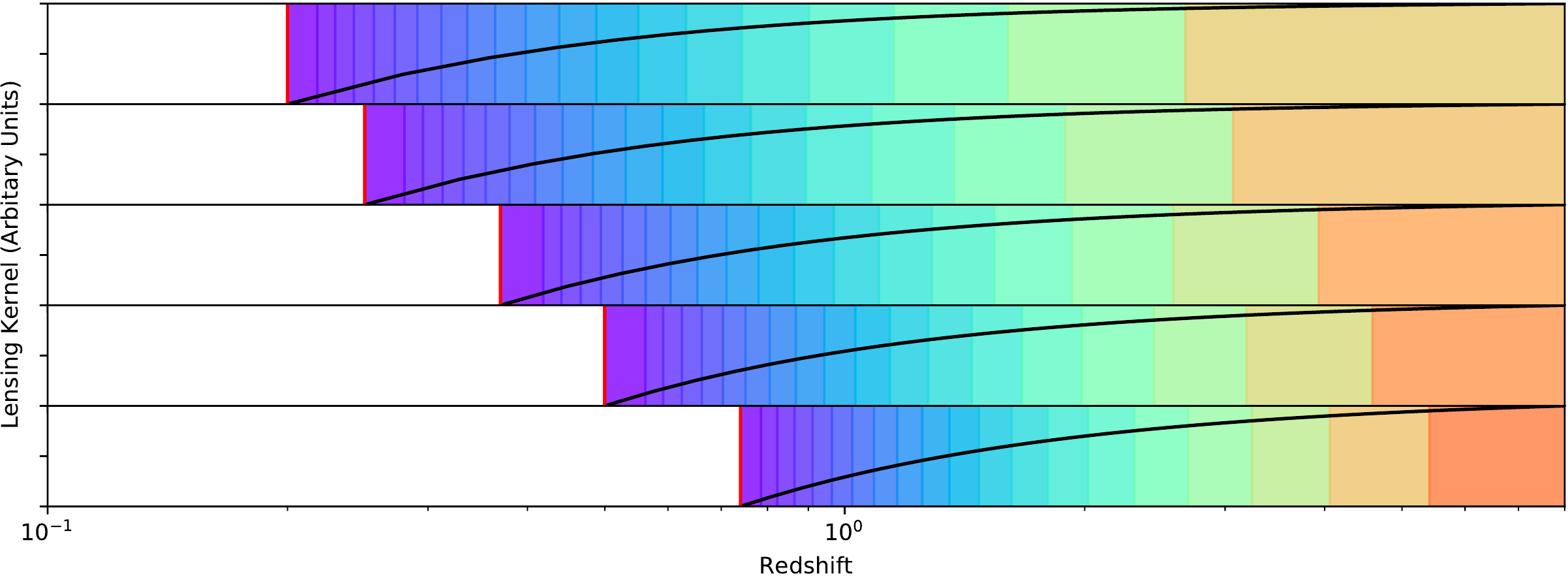}
\caption{\label{fig:lensSourceConfig} {\it The lens-source configurations of the simulations}. For each simulated halo we extract five redshift slices: $\zl=(0.20, 0.25, 0.37, 0.5, 0.74)$, shown as each row here. For each lens we select 21 source redshift slices to integrate over, each one equally spaced in change in lensing kernel (i.e. $\Delta [\dls\dl/\ds]$) up to $\zs=8$. }
\efigs
\begin{table*} 
\caption{The statistics of each simulated halo.  We show the ID of the halo (both the ID from \citet{oppenheimer16} and \citet{lensingEAGLE}), the halo mass (total mass inside a radius where the density is $200$ times the critical density at the given redshift), the virial radius, the stellar mass of the central galaxy, the consequential velocity dispersion from the fitting function in \citet{velocityStellarMassRelation}. The final two columns give the  stellar effective radius in 3D and the 2D average over many projections using the stellar particles belonging to the main galaxy within 300 (100) kpc. } 
\label{tab:haloStats} 
\begin{center}
\begin{tabular}{ccccccc} 
\hline 
ID & $M_{\text{200c}}$ &  $r_{\text{200c}}$ & $M_{*}$  &$\sigma$ & $r_{*,e}$(3D) &$\langle r_{*,e}(2D)\rangle$\\
\hline
& [$\rm{M_{\odot}}$] &[kpc] & [$\rm{M_{\odot}}$] & km/s & [kpc] & [kpc] \\
\hline 
B008 / 1 & 1.06 $10^{13}$ &401 & 9.45 $10^{10}$  &192 &12.76 (8.45) & 9.31 (8.12)\\
B009 / 2& 1.05 $10^{13}$ &401 & 1.00 $10^{11}$  & 194 & 22.31 (13.31)& 27.11 (13.93)\\
B005 / 3& 6.40 $10^{12}$ &340 &5.07 $10^{10}$   & 159 & 6.24 (4.64) & 5.80 (4.24)\\
B002 / 4& 3.99 $10^{12}$  &290 &5.48 $10^{10}$  & 163 &6.46 (5.09) & 5.15 (4.09)\\
\hline

\end{tabular}

\end{center}
\end{table*}

\subsection{Micro-lensing}
In a similar way we estimate the impact of micro-lensing on the distribution. \cite{microlensingTimeDelay} calculated the cumulative probability distribution of the micro-lensing induced time delays for the lenses RXJ1131-231 and HE 0435-1223. They used numerical simulations to shoot light rays through a de Vaucouleurs model of the stellar distribution to calculate a time delay map. Taking 3000 different sight lines they found that the mean delay was zero for a AGN disk with no inclination, and of order $\sim 1$day for inclined disks. Moreover, the variance in the time delay is of order days. Therefore we model the impact of micro-lensing by a Gaussian with a zero mean and width of $1$ day and $2$ days.  This is a conservative estimate since  \cite{microlensingTimeDelay} investigate the impact of quadruply imaged quasars where projected stellar density is higher and hence will results in large micro-lensing.  Given that this affects each time delay it is simply a convolution with the intrinsic distribution. Figure \ref{fig:LoS} shows the resulting distribution. We find that for a reasonable delay, micro-lensing could have up to a $10\%$ impact on a single lens-source configuration. We note here that although milli-lensing by substructure in the lens can have impact on the time delay, it is expected to be an order of magnitude lower than micro-lensing considered here \citep{timeDelayMiliLensing}, and therefore we do not consider this systematic.

\subsection{Combining different lenses and source planes}

Now with a full estimate of a single lens-source configuration we want to set up a framework in which we can integrate this over a lens and source population. In order to do this we must weight each lens and source by the volume of the redshift in question, consider that sources behind lenses will be magnified and that the number density of quasars at different redshifts changes. In summary the final estimated PDF will be,
\be
p(\log(\Delta t)) =  \int d\zs \int d\zl   \int dM   \int dM_{v}~p(\zl, \zs, L, \mu,M_{v}),
\label{eqn:integrateAll}
\ee
where $p$ is the probability distribution of a single, source-lens configuration, given by
\be
p = \frac{dV}{d\zs}  \frac{d\Phi}{dM} \frac{dV}{d\zl} \frac{dN}{dM_{v}}\sigma_{\rm lens} p(\log(\Delta t_O)),
\ee
where the final term is the number of time delays in an interval $d\log(\Delta t)$ for a given lens with mass $M_{v}$,at a given lens redshift, $\zl$, for a source distribution (or luminosity function) $d\Phi/dL$ (and magnitude $M$) at a redshift, $\zs$, the volume factor for the lens and source is given by
\be
\frac{dV}{dz_i} = \frac{c dt}{dz_i}(1+z_i)^3,
\ee
and $\sigma_{\rm lens}$ is the ``magnification bias'' caused by the magnification induced by the foreground lens increasing the source number density by bringing faint sources in from beyond the magnitude limit of the survey and is given by
\be
\sigma_{\rm lens} = \int \frac{d\mu}{\mu}\frac{d\Phi/dL(L/\mu)}{d\Phi/dL(L)}, \label{eqn:magBias}
\ee
where we adopt the analytical quasar luminosity function from \cite{quasarLumFunc},
\be
\frac{d \Phi}{dL} = \frac{\Phi^\star}{10^{0.4(\alpha(z)+1)(M-M_\star(z))} + 10^{0.4(\beta(z)+1)(M-M_\star(z))}},
\ee
assuming a magnitude limit of $M_{AB}=27$ \citep{LSST}, $\alpha=-3.23$, $\beta=-1.35$ and $\log(\Phi^\star)=a + bz +cz^2$ with $a = -6.0991$, $b = 0.0209$ and, $c = 0.0171$. The final cumulative distribution function is therefore
\be
p(> \log(\Delta t)) = 1 - \int_{-\infty}^{\log \Delta t} p(\log(\Delta t)) d \log(\Delta t).
\ee

\fig
\includegraphics[width=0.49\textwidth]{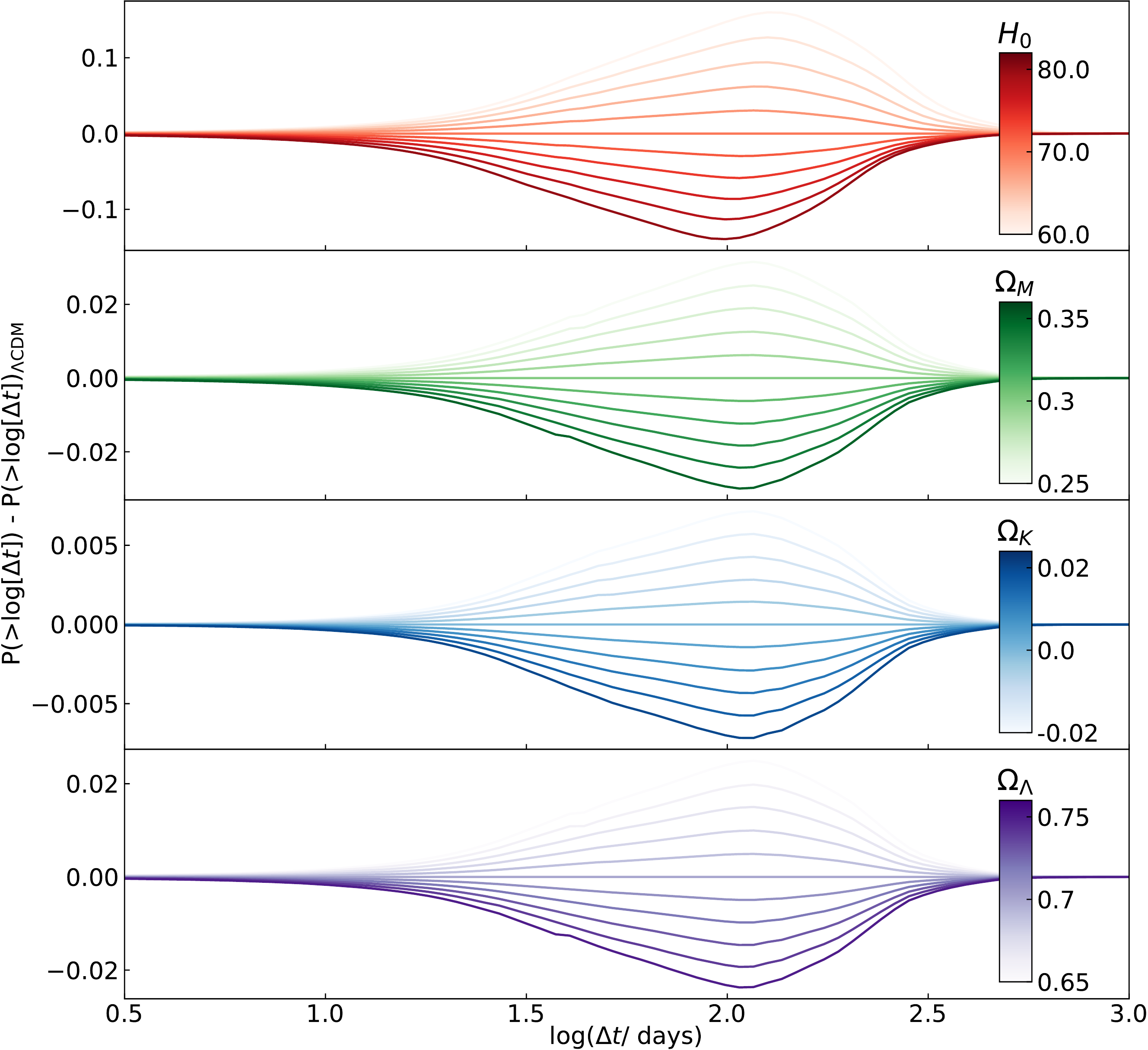}
\caption{\label{fig:cosmoDep} {\it Cosmological dependence of the CDFs}. Each panel shows the change in the CDF relative to $H_0=70$km/s/Mpc, $\Omega_{\rm M} =0.3$, $\Omega_{\rm K} =0.$ and $\Omega_{\rm \Lambda} =0.7$. Although the ranges are relatively arbitrary, we see that the dominant parameter is $H_0$ with changes of $\tilde10\%$.}
\efig

\subsection{Application to hydrodynamical simulations}\label{sec:nbody}
Now we have set up the framework in which to extract time delays and fold in line-sight structures and micro-lensing, we move beyond simple analytical profiles. We adopt a suite of {\it n}-body simulations that uses a full hydrodynamical prescription that includes the EAGLE baryonic prescription and a modified {\sc GADGET 3} code \citep{eagle}. The suite consists of four halos that are zoomed in simulations of giant ellipticals from a larger simulation, identified as the most appropriate for re-simulation as determined by \cite{oppenheimer16} and suitable for lensing observations by \cite{despali17b}. A detailed description of these simulations can be found in \cite{lensingEAGLE}, however here we present a concise summary. The simulations include pressure-entropy smooth particle hydrodynamics \citep{crain15,eagle}, stellar evolution, supernova feedback, active galactic nuclei and cooling. Haloes were identified using a Friends of Friends algorithm and their properties using {\sc Subfind} \citep{springel01,dolag09}. Each of the four volumes have gas particle mass of $m_{\rm gas} = 2.3\times10^{5}M_\odot$ with cosmological parameters consistent with the \cite{planckParsFinal} constraints: $h_{0}=0.6777$, $\Omega_{0}=0.307$, $\Omega_{\rm b}=0.04825$, $\Omega_{\Lambda}=0.693$, $n_{\rm s}=0.9611$ and  $\sigma_8=0.8288$. Table \ref{tab:haloStats} shows the statistics of each halo simulated, including the group ID (ID from \cite{oppenheimer16} and \cite{lensingEAGLE}), the halo mass, the virial radius, the stellar mass, the estimated velocity dispersion which we assume follows the relation in  \cite{velocityStellarMassRelation}, the effective stellar radius of the 3D and 2D distribution of stellar particles.  In order to carry out the lensing analysis we first create two-dimensional mass maps of each simulated halo at various redshifts. To do this we extract cones from five redshift snapshots ($\zl=0.20, 0.25, 0.37, 0.50, 0.74$) and create 2 dimensional mass maps projected in all three dimensions (i.e. 15 projected mass maps per halo), each out to an x-y distance of $100$ pkpc (physical kpc) and with a projected depth of $1$pMpc, and a pixel resolution of $0.1$pkpc. A projected depth of $1$pMpc should account for the contribution to environmental effects from all nearby structures $M\lesssim10^{14}M_\odot$.  We extract three key features from the simulations that will be important in the analysis: {\bf 1.)} The number of substructure inside a projected $20$kpc (and $1$Mpc deep) cylinder, with a mass $M_{200c}>10^{7}M_\odot$; the density profile derived from the projected inner $10$kpc of the galaxy (and then de-projected); and the total {\it projected} mass within $5$kpc of the galaxy.

\figs
\includegraphics[width=0.49\textwidth]{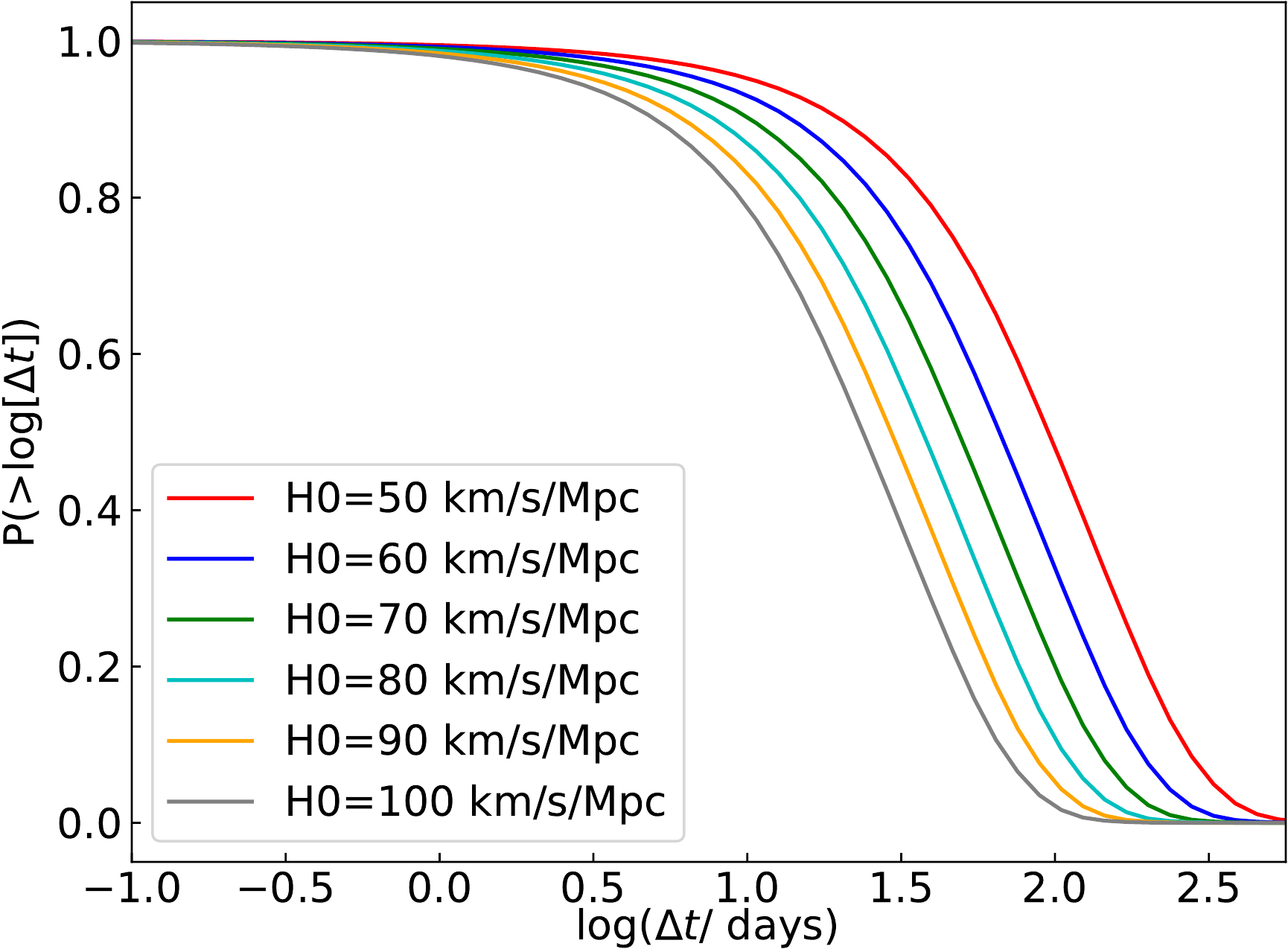}
\includegraphics[width=0.49\textwidth]{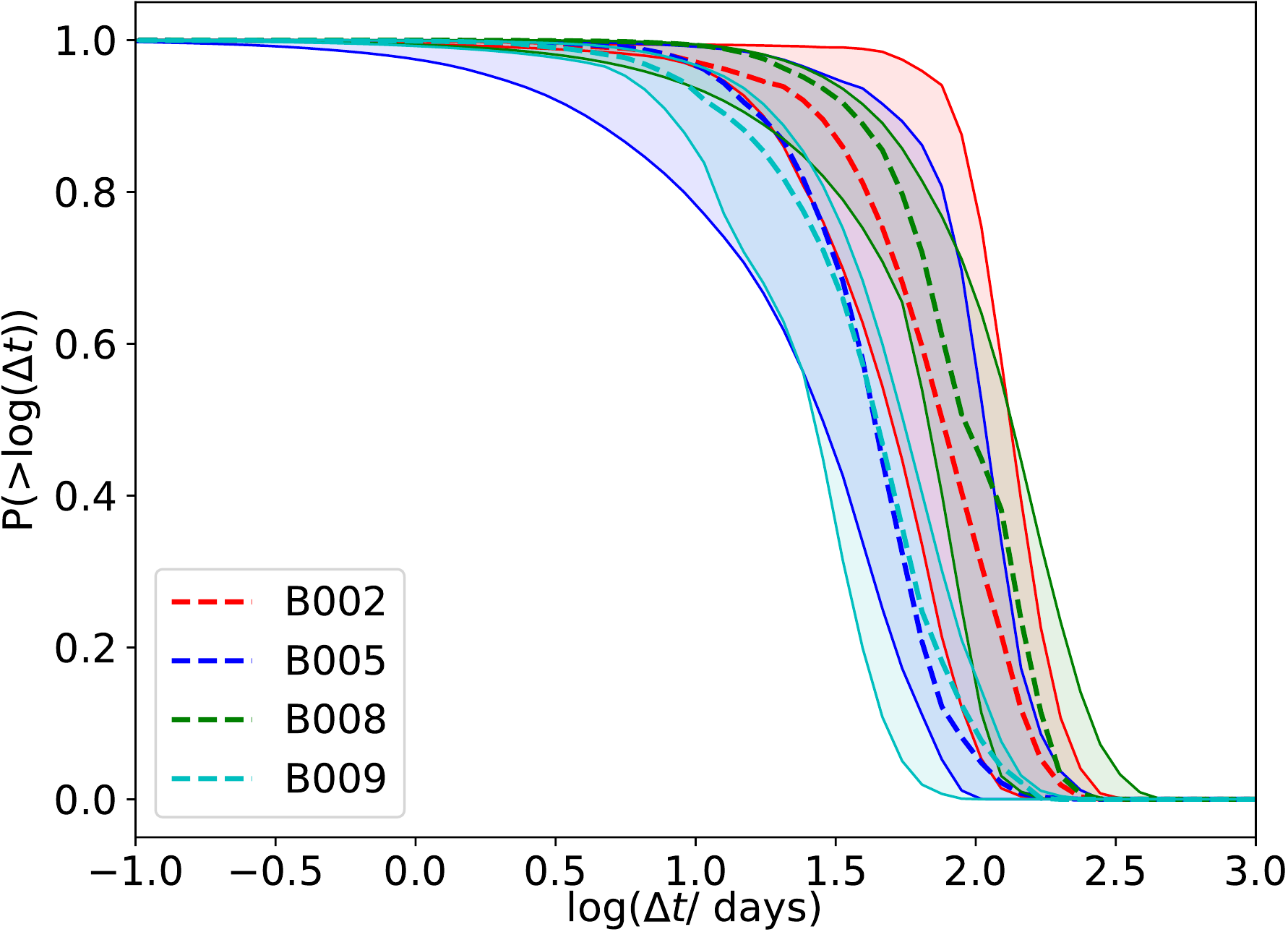}
\caption{\label{fig:allLenses}  {\it Final CDFs as predicted by the simulations}. {\it Left:} The cumulative density function for all lenses for different values of $H_0$, including line-of-sight effects, micro-lensing and magnification bias. {\it Right: } The halo to halo variance for all field in the four volumes. The dashed line gives the median value and the shaded regions contain the 16\% and 84\% percentile of CDFs.}
\efigs

Equation \eqref{eqn:integrateAll} implies that we can integrate to an infinitely small $d\zl$ and $d\zs$, however in practice with finite amount of hard-disk space and computation power this is unfeasible. From the 60 extracted lens planes (four halos, three projections, five redshift slices) we follow the prescription laid out in \ref{sec:theory} and calculate the probability density function $p_{\rm T}(\Delta \log t)$ for a single source-lens configuration. For a given source plane we then convolve with equation \eqref{eqn:los} to find $p_{\rm O}(\Delta \log t)$. We then calculate the magnification bias, $\sigma_{\rm lens}$ from equation \eqref{eqn:magBias} for each multiple image given a quasar luminosity function at the given redshift (assuming a limiting magnitude of $m_{\rm AB, limit}=27$) \citep{observingStrategyLSST}. The magnification value we choose in equation  \eqref{eqn:magBias} will depend on the observing strategy. Here we follow \cite{lensedQuasarsForLSST} and choose the magnification of the fainter image. From this we now have our final probability density function (for a single lens-source configuration). We then do the same for a range of source redshift planes choosing equal steps in the lensing kernel (i.e. $\Delta \left[ \dls\dl/\ds\right]$)  from the lens redshift to a maximum source redshift of 8, with 21 redshift bins. Figure \ref{fig:lensSourceConfig} shows the lens-source configuration for the five lens redshifts we use. Each colour shows a source redshift bin, the solid black line shows the lensing kernel of each redshift with respect to the lens redshift given by the solid red vertical lines.

\subsection{Implementing cosmological dependence}
We now extend the CDFs of the cosmological simulations at a single cosmology to {\it any} cosmology. To completely encapsulate the impact of a change in cosmology on the CDF we would need to re-simulate the entire cosmological box with varying cosmological parameters. However, this is currently unfeasible, moreover, the dominant modification will be to the time delay distance and the weighting of each redshift slice by the volume. Hence, to simulate a CDF for a given set of cosmological parameters, we re-calculate equation \eqref{eqn:integrateAll}, with the new cosmology and note that future works should study the impact on structure formation that this method does not encapsulate. Figure \ref{fig:cosmoDep} shows multiple examples of this re-calculation showing the total CDF relative to a $\Lambda CDM$ CDF (i.e. $H_0=70$km/s/Mpc, $\Omega_{\rm M} =0.3$, $\Omega_{\rm K} =0.$ and $\Omega_{\rm \Lambda} =0.7$) . We find that as expected, the time delay CDF is most sensitive to deviations in $H_0$, with curvature having the smallest. Given the changes in the CDF, we do not expect competitive constraints on anything except the Hubble Parameter.

\section{Results}\label{sec:results}

We now present the full CDF, including all sources of systematics, integrating over all lens and source configurations. The left hand panel of Figure \ref{fig:allLenses} shows the cumulative probability of observing a time delay, i.e. $p(> \log(\Delta t)$ for six Hubble parameters. We find that the change in Hubble constant acts to simply shift the CDF along the time-delay axis.

Following the study of the ensemble lenses, we investigate the halo-to-halo variance. The right hand panel of Figure \ref{fig:allLenses} shows the results of this test. We show the median value of each halo with a dashed line, and the shaded regions show the 16\% and 84\% of all CDFs. We see that the halo to halo variation is quite large, with B002 and B008 exhibiting higher expected time delay distributions than B005 and B009. This is much more interesting if you consider that B002 has a lower velocity dispersion than B009. However, as show in  
 \cite{lensingEAGLE}, B009 has significantly more substructure than B002, by a factor of four. We therefore follow this up by studying three key properties of the lens, the number of substructures, the halo total mass density profile and the total mass within $5$kpc, which roughly translates to the Einstein Radius of these lenses. Figure \ref{fig:halo2haloVar} shows the results of these three tests.
 
 In the left hand panel of Figure \ref{fig:halo2haloVar} we show the expected median time delay ($\log(\Delta t_{\rm med})$) in black and the most likely time delay ($\log(\Delta t_{\rm max})$) in red as  function of number of structures inside $20$ kpc and with a mass $M/M_\odot>10^7$. We see that both the most likely and median time delay are effected by the amount of substructure in the lens, acting to reduce it. This is interesting given that \cite{substructuresTimeDelay} found that ignoring substructures did not bias the estimated Hubble constant from quadruply imaged quasars. However, this study looked  looked at the impact of substructures on specific lenses. We see in the left hand panel of Figure \ref{fig:halo2haloVar} that although the general trend is for an increase in the number of substructures tends to decrease the expected time delay, there is a large amount of scatter and it is not unlikely that a single lens system with many substructures will have an increased expected time delay. As such comparing the {\it statistical} impact of substructures with the impact of substructures of a specific lens configuration is difficult and hence this is not comparable with the study by \cite{substructuresTimeDelay}.

\figs
\includegraphics[width=0.303\textwidth]{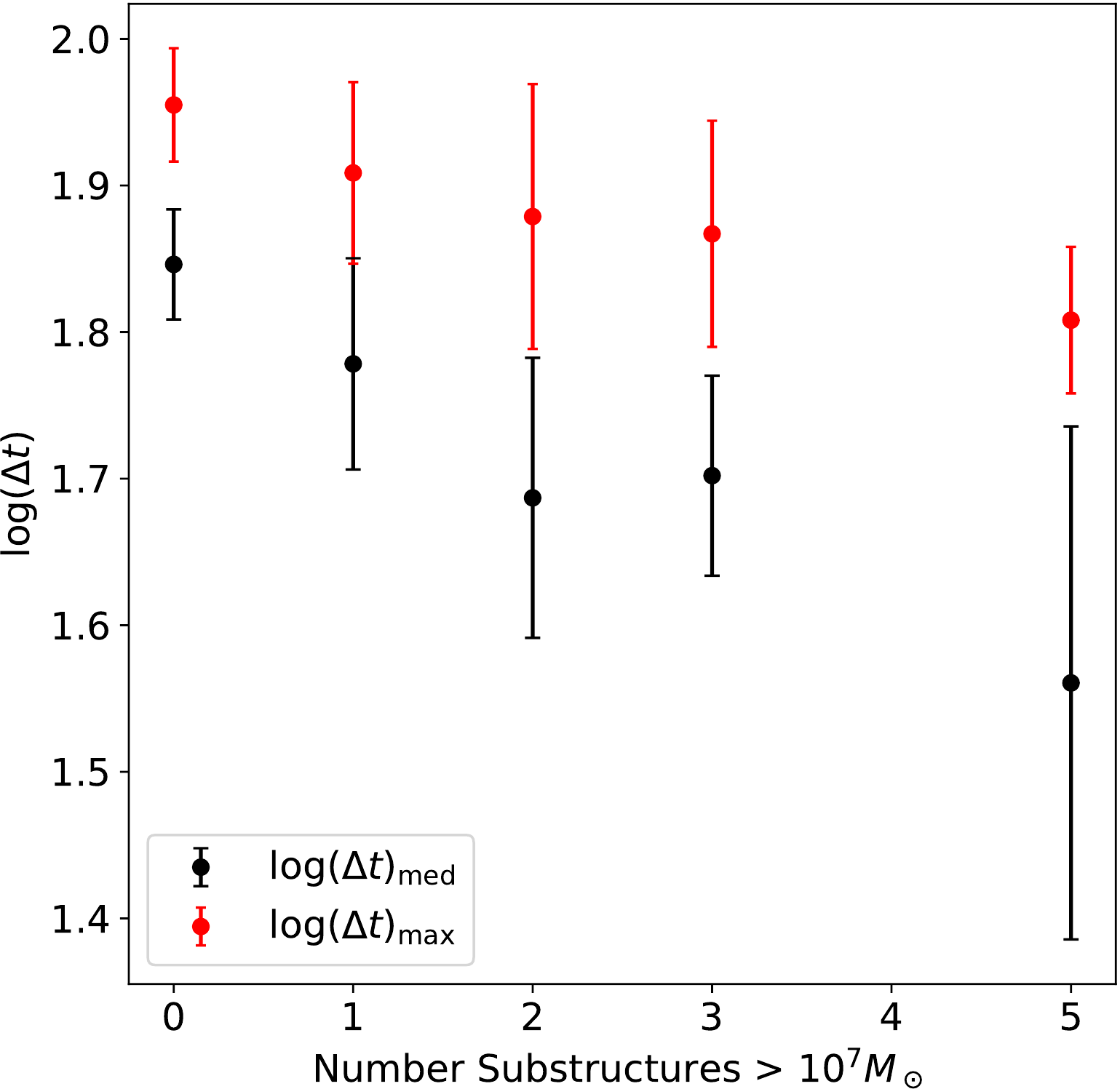}
\includegraphics[width=0.304\textwidth]{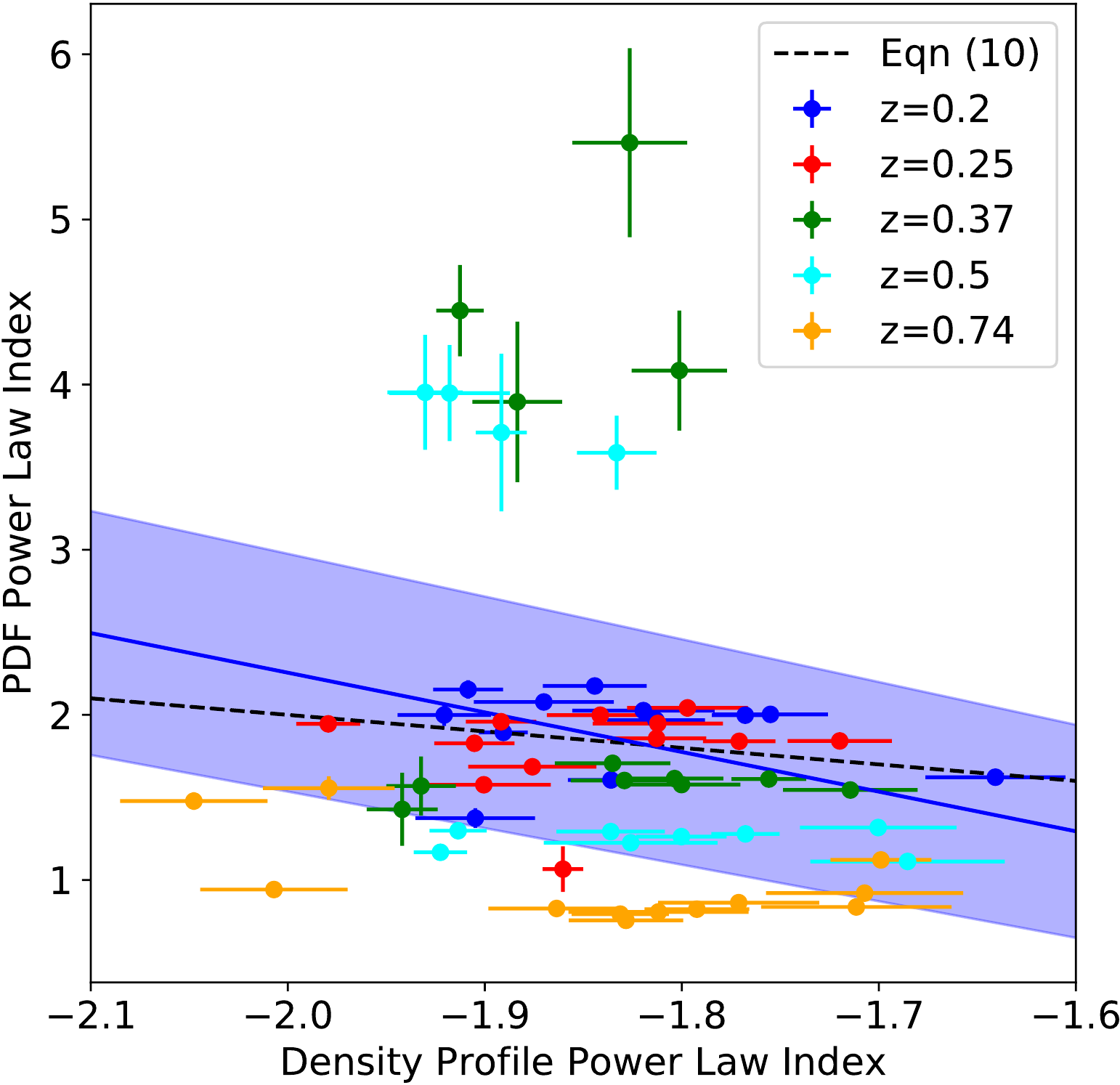}
\includegraphics[width=0.32\textwidth]{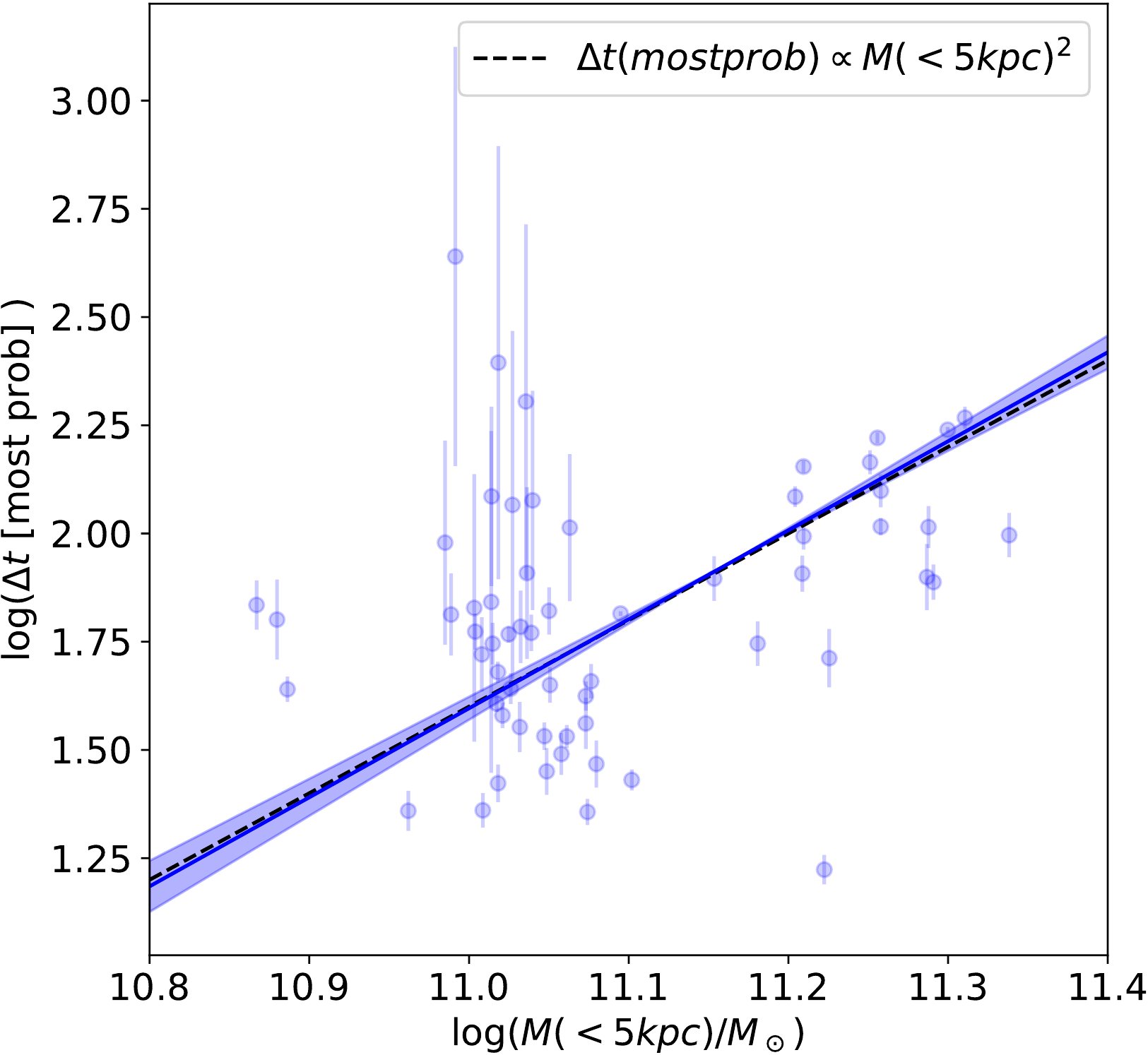}
\caption{\label{fig:halo2haloVar}  {\it Left:} The median (black) and most likely (red) time delay as a function of number of substructures inside $20$kpc and with a mass $M/M_\odot>10^7$. {\it Centre:} The relationship between the power law index and as a function of density profile power law index and lens redshift. We see the clear degeneracy between both parameters. {\it Right: } The relationship between the most probable time delay (i.e. the peak of the PDF) and the total mass within $5$kpc of the centre of the galaxy.}
\efigs

The central panel explores the impact of the varying density profile index, testing the relation stated in equation \eqref{eqn:logT} that the power law of the PDF correlates directly with the power law of the halo. We test this by first measuring the projected density profile of each individual lens and de-project assuming spherical symmetry.
We then fit a power law model to each PDF, fitting only to the $p(\log\Delta t)>10^{-2}$. The central panel of Figure \ref{fig:halo2haloVar} shows the measured power law index of the PDF as a function of the density profile, for each halo, with the redshift of the halo in colour. The blue line shows the fitted trend with the black dashed line showing the expectation from equation (10). We see that the mean relation matches that expected, however we notice a larger variance in the measured PDF, with a clear trend in redshift. We also note that the points at $z=0.37$ have larger amounts of substructure in the lens and therefore mean the power law fit is not a good one. 

Finally we test how the amount of projected mass within $5$kpc of the lens effects the most probable time delay (i.e. the peak of the PDF). We show the measured values from each halo in the simulations with the fitted line and error in blue. We show the $\Delta t_{\rm prob}\propto M^2$ in the dashed line, which is what is expected from an SIS. We find that these match extremely well.

To summarise we find that there exists a large halo-to-halo variance, with substructures in the lens shifting the expected time delay to lower values, and although the lenses follow the expected relation, the density profiles seem to be shallower than $\alpha=-2$ (that had been simulated in previous studies), however the most probable time delay closely follows a $\Delta t\propto M^2$ relation, which is naively what we would expect.

\fig
\includegraphics[width=0.5\textwidth]{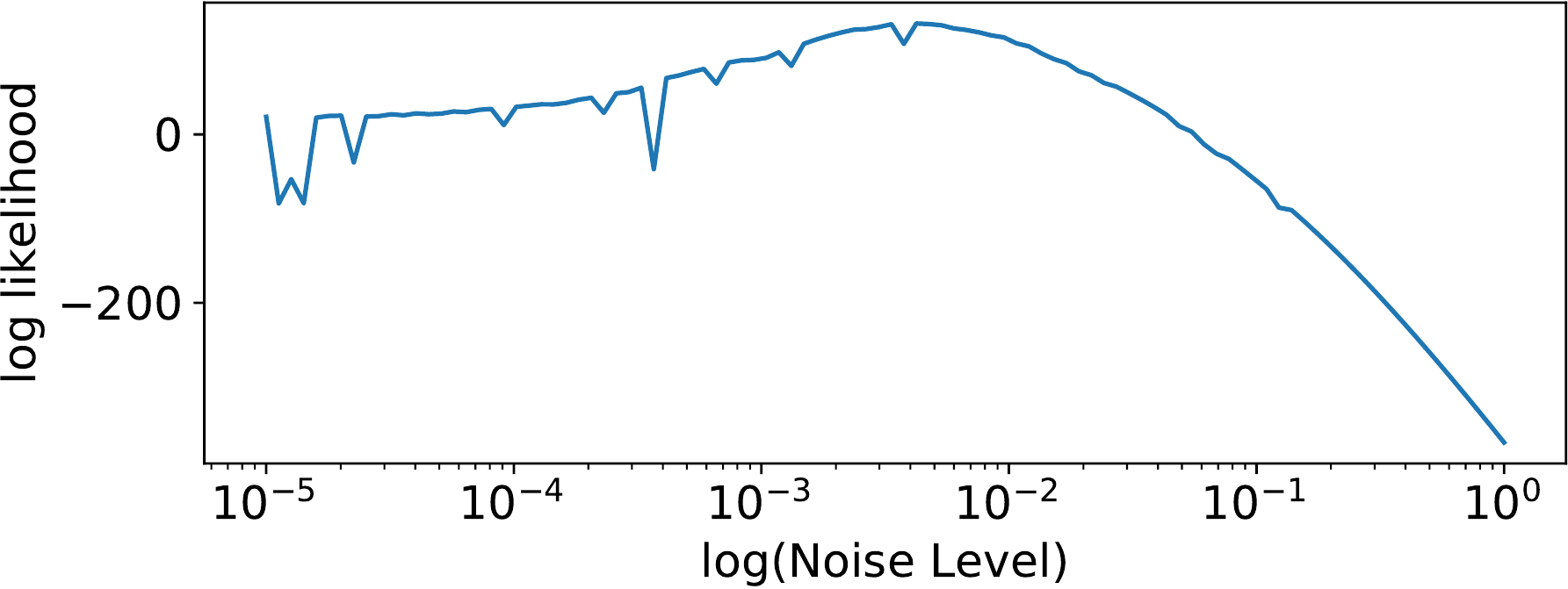}
\includegraphics[width=0.5\textwidth]{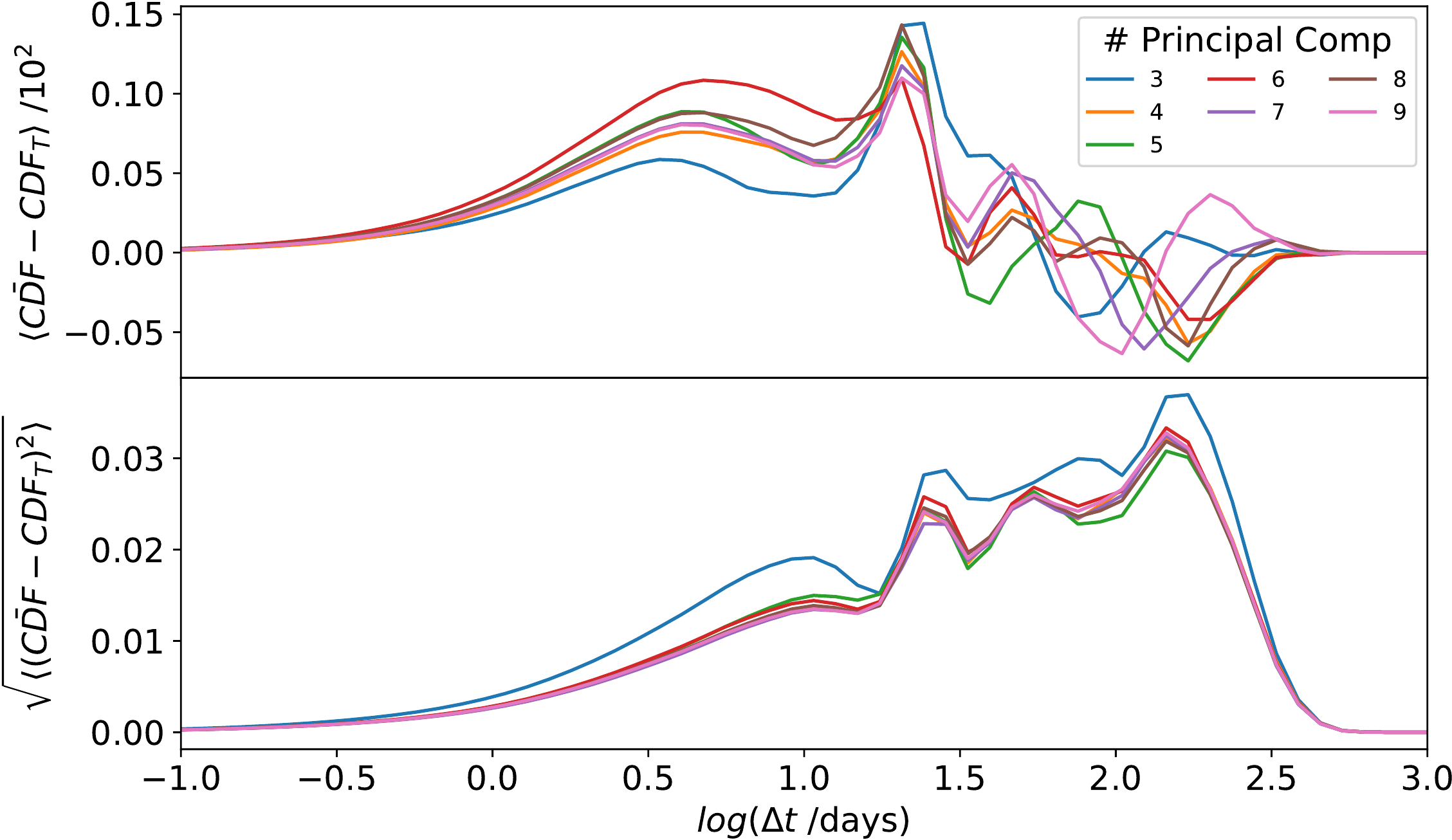}
\caption{\label{fig:pcaAnalysis} {\it Top:} Choice of noise level for the Matern Kernel in the Gaussian Processor Regressor (GPR). {\it Bottom:} GPR as a function of number of principal components used. The top panel shows the mean difference between the estimated CDF ($\bar{CDF}$) and the true CDF for different number of principal components (different colours). We find that the GPR can predict the CDF to within 0.1\%  bias.  The bottom panel shows the root mean square of the GPR for different number of principal components. We find that the statistical variance is $\sim2\%$.  }
\efig
\subsection{Constraints on current data}\label{sec:constraints}

We now look to data to see how precisely we can measure $H_0$. However, in order to do this, we must construct a model to compare to the data. To do this we carry out the following procedure,
\begin{enumerate}
\item Reduce the dimensionality of each halo's CDF using a principal component analysis (PCA);
\item For a given set of features, ($\zl$, $\alpha$ (density profile power law index)) and $\mfive$, train a Gaussian Process Regressor that can predict the principal components for the fiducial cosmology;
\item Learn the expected shift in the CDF due to change in the cosmological parameter set (i.e. Figure \ref{fig:cosmoDep}), by fitting a simple linear regressor;
\item With an algorithm that can now predict the CDF for a given lens redshift, power law index and cosmological parameter set, estimate the parameters of any CDF;
\item In a Bayesian framework, use an Monte Carlo Markov Chain (MCMC) to estimate the parameters from a mock observation of time delays.
\item Apply the framework to observed data.
\end{enumerate}

\subsubsection{PCA Analysis to reduce the dimensionality} 
We first reduce the dimensionality of the CDFs by decomposing them in to their principal components with a principal components analysis (PCA). PCA analysis is a popular way to compress information, often in, for example, image analysis. PCA is founded on the idea that a data vector will have a number of principal, orthogonal axes, such that the high dimensionality of a data vector can be reduced. By linearly combining each component of the data vector with some weight we are able to transform the data in to a subspace that has the highest possible variance, i.e. it explains the largest amount of the data. This amounts to the first principal component. This component can then be removed and the subsequent principal components can be found. 

We carry out a PCA analysis on each individual time delay CDF. In total we have 60 CDFs  (four halos, three projections, and five lens redshifts, $\zl$). We adopt the PCA analysis from the python package {\sc sckit-learn}\footnote{\url{https://scikit-learn.org/stable/modules/generated/sklearn.decomposition.PCA.html}} that follows a probabilistic PCA\footnote{\url{http://www.miketipping.com/papers/met-mppca.pdf}}. 

\subsubsection{Gaussian Process Regressor to predict the principal components}
Now with our principal components at discrete regions in the parameter space (for a fiducial cosmology) we look to create a predictive model that can interpolate between these and predict at all regions. To do this we adopt a Gaussian Process Regression technique (GPR). A GPR is a supervised machine learning method that adopts a  mixture of Gaussian distributed functions that allow us to interpolate between variables by mapping a function or kernel to an N-dimensional space. GPRs have been used to constrain the Hubble constant in the past, for example \cite{LiaoStatHubble} combined strong gravitational time delays and Supernova to estimate $H_0=72.8^{+1.6}_{-1.7}$km/s/Mpc, constituting a 2.3\% uncertainty on the Hubble constant. For more about GPR please see \cite{GPR}.

\figs
\includegraphics[width=\textwidth]{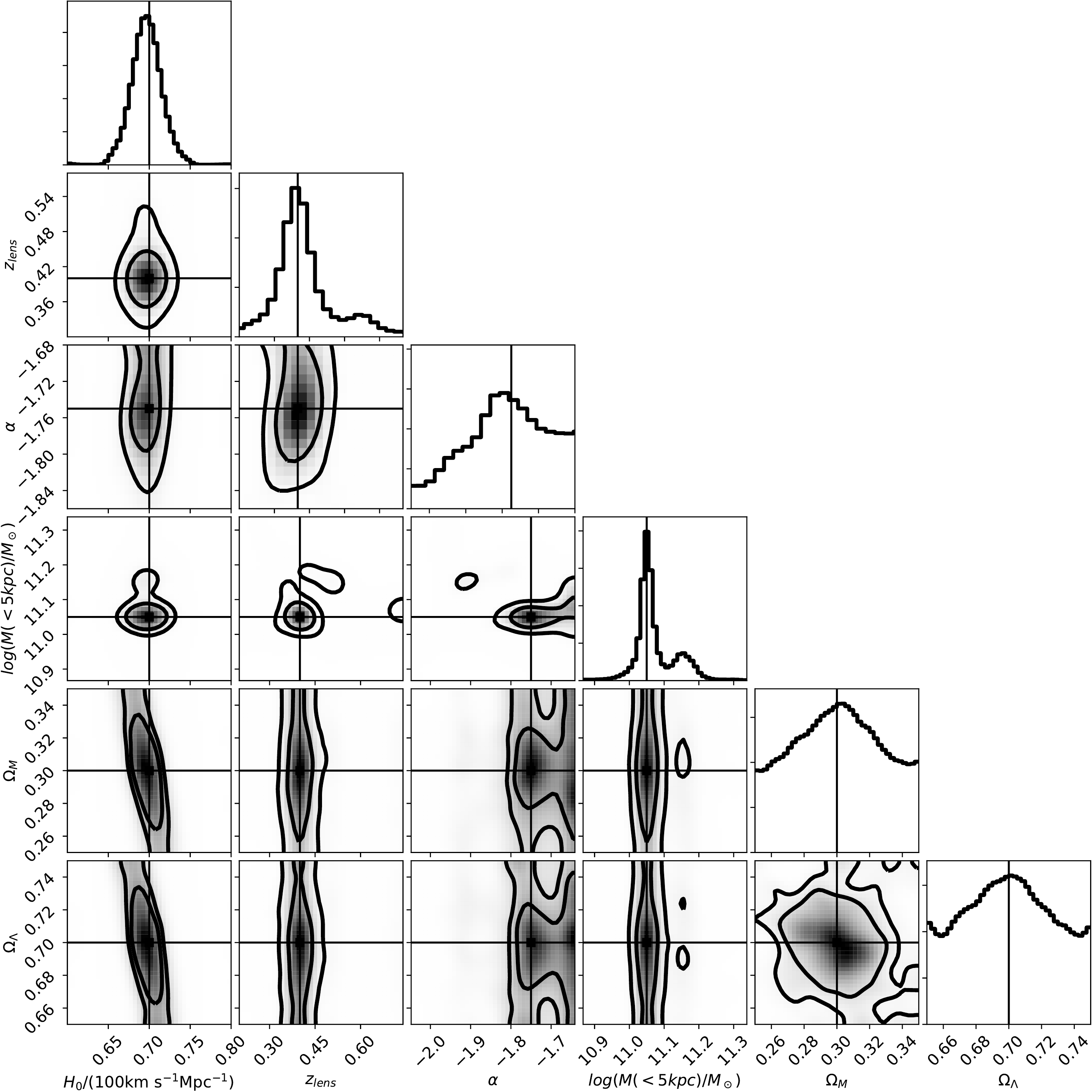} 
\caption{\label{fig:consistencyCheck} Consistency check on the Gaussian Process parameter estimation. We generate a mock sample of $n_{\rm lens}=10^3$ observed quasars from the trained model and then using an MCMC to find the best fit parameters. here we show the $0.5$ and $1-\sigma$ contours with the black marks showing the true value. }
\efigs
In order to apply the GPR to this problem we fit and train on each principal component with input features $\Theta=(\zl, \alpha,\mfive)$ (i.e. lens redshift, density profile power law index and the mass within $5$ kpc), this way for each component the GPR learns the relationship between these features and the target principal component value. In order to train the most accurate GPR we carry out a MCMC over the meta-variables of each kernel available in the scikit-learn package\footnote{\url{https://scikit-learn.org/stable/modules/gaussian_process.html}}. We find that the combination of the Matern kernel\footnote{\url{https://scikit-learn.org/stable/modules/generated/sklearn.gaussian_process.kernels.Matern.html\#sklearn.gaussian_process.kernels.Matern}} returns the highest log-likelihood value during the fitting process. The choice of meta-parameters is important in this situation since they will govern how well the GPR is at predicting the PCA. We initially adopt a length scale=1. and  $\nu=3/2$ and find that the log-likelihood of the GPR is insensitive to variations in these parameters, however the value of the noise floor, $\alpha_{\rm N}$ does alter the log-likelihood. We therefore carry out a simple search for the noise value that returns the highest log-likelihood. The top panel of Figure \ref{fig:pcaAnalysis} shows that the value of $\alpha_{\rm N}$ with the highest log likelihood is $\alpha_{\rm N}=4\times10^{-3}$. 

We now extend this algorithm that can predict the CDF for a given set of physical parameters, to include a complete set of cosmological parameters. To do this, we calculate the shift in the CDF due to a change in the cosmological parameter set by first setting up a grid of cosmological parameters: $H_0=\left[60, 80\right]$, $\Omega_M=\left[0.25, 0.35\right]$,
$\Omega_\Lambda=\left[0.65, 0.75\right]$, and $\Omega_{\rm K}=\left[-0.02, 0.02\right]$, with each having five equally spaced intervals except $H_0$, which has eleven. We calculate the CDF at each point in the 4-dimensional parameter grid and the associated cosmological shift and then fit a linear regression to this shift\footnote{\url{https://scikit-learn.org/stable/modules/generated/sklearn.linear_model.LinearRegression.html}} such that we can estimate the shift at {\it any cosmology}. We now have a algorithm that can estimate the CDF given 7 parameters, $\Theta=(\zl, \alpha,\mfive), H_0, \Omega_M, \Omega_\Lambda, \Omega_K)$

We show the accuracy of the GPR in Figure \ref{fig:pcaAnalysis}. The top panel shows the mean difference between the predicted CDF and the true CDF. We see that the systematic bias has a maximum of $<0.1\%$ for all principal components above three. The bottom panel shows the root mean square between the true and the predicted CDF. We find above five principal components there is no discernible difference. As such we decide to use 6 principal components going forward. Moreover we find that the intrinsic dispersion in the estimator is $\sigma_{\rm GPR}\sim1.5\%$, signifying the precision limit of this model.

\figs
 \includegraphics[width=\textwidth]{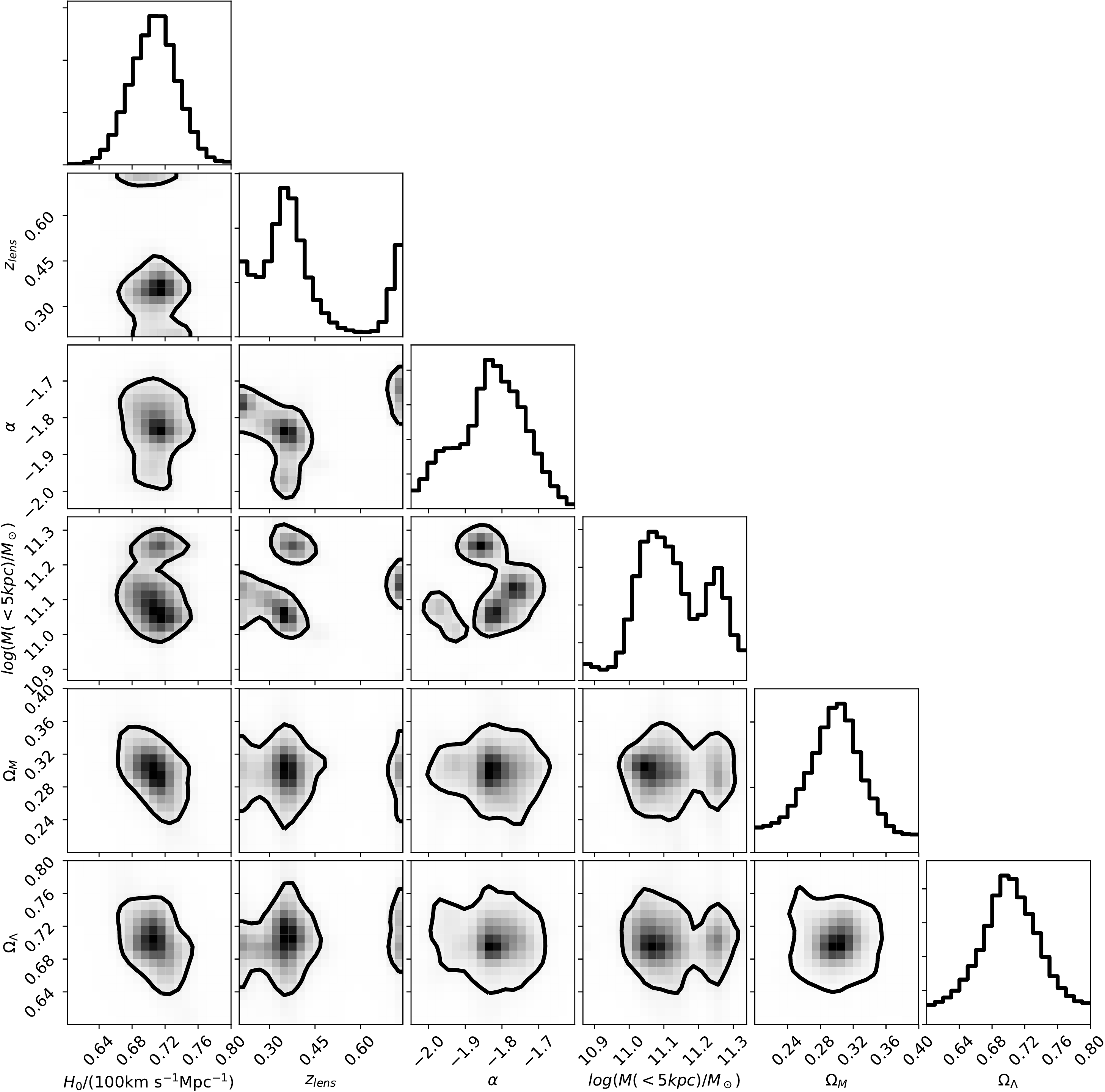}
\caption{\label{fig:constraints} The posteriors from our final fit. Contours show the $1-\sigma$ error. We run three different fits, one with a fixed cosmology (i.e. $\Omega_{\rm M}=0.3$, $\Omega_{\rm \Lambda}=0.7$, $\Omega_{\rm K}=0.$), one assuming a flat Universe and one completely free. Here we show the results assuming a flat Universe, the other two fits can be found in Appendix \ref{sec:finalResults}.  We find  $H_0=70\pm3$km/s/Mpc, $\zl = 0.36_{-0.08}^{+0.3} $, $\alpha=-1.8_{-0.1}^{+0.08}$,   $\mfive=11_{-0.08}^{+0.1} $, $\Omega_{\rm M} = 0.3_{-0.05}^{+0.05} $ and $\Omega_{\rm \Lambda}=0.7_{-0.05}^{+0.05}$. }
\efigs

\subsubsection{Validating the model and sample}

Now armed with an algorithm that can predict the CDF for any given $\Theta=(\zl, \alpha,\mfive), H_0, \Omega_M, \Omega_\Lambda, \Omega_K)$ we can estimate $H_0$ from a given sample of time delays. We first test the self-consistency of our algorithm. We simulate a mock CDF with $\Theta=\{H_0=0.7,~\zl=0.4,~\alpha=-1.75,~\mfive = 11.05,\Omega_{\rm M}=0.3,\Omega_{\rm \Lambda}=0.7\}$ and a fixed value of $\Omega_{\rm K}=0$. We simulate a survey of $n_{\rm lens}=10^3$ observed lenses, 100 times. We fit our model to the data in the same flat prior space that we laid out our grid of cosmological parameters, i.e.  $H_0/(100$km/s/Mpc)$=\{0.6,0.8\}$, $\zl=\{0.,0.74\}$, $\alpha=\{-2.1,-1.4\}$,  $\Omega_M=\{0.25, 0.35\}$ and $\Omega_\Lambda=\{0.65, 0.75\}$. At each step of the MCMC we calculate the expected CDF from the GPR and compare to the observed via the Cram\'er-von Mises criterion that is the sum of square of the distance between the data, $D$, and the model $M$.
\be
p(m(\Theta)|D) = \sum(CDF_{\rm D} - CDF_{\rm M}(\Theta))^2.
\ee
We use the publicly available package {\sc emcee}\footnote{\url{https://emcee.readthedocs.io/en/stable/user/sampler/\#}}, using a burn in length of $n_{\rm burn}=500$ and a sampling chain length of $n_{\rm chain}=1000$. Figure \ref{fig:consistencyCheck} shows the results of this test. We find that we return the input $H_0$, $\zl$, density profile index $\alpha$ and the two cosmological parameters, $\Omega_{\rm M}$ and $\Omega_{\rm \Lambda}$. Interestingly we find that the posterior of the mass within $5$ kpc has a degeneracy with $\alpha$, whereby, the sampler does return the input value, but it is also possible to produce similar features  in the CDF with a steeper density profile and less mass within $5$kpc.

\subsubsection{Constraints from current data}
Now with our self-validated code we fit our model to the current observed data. We adopt time delays with their associated errors from \cite{allTimeDelays} (and references therein). Table \ref{tab:timeDelays} in Appendix \ref{sec:timeDelays}  gives an overview of the each object used, reference for the measured time delay, the lens redshift, source redshift and the estimated time delay and error. We create a CDF from these time delays and fit our model. To incorporate the error bars in the estimates, we then Monte Carlo the CDF 100 times, each time resampling the CDF from the stated error bars and refitting the model. We carry out this procedure in three different cosmologies, the first, and our stated constraints, a flat Universe, assuming $\Omega_{\rm K}=0$, the second not assuming a flat Universe and the third a fixed cosmology ($\Omega_{\rm M}=0.3$, $\Omega_{\rm \Lambda}=0.7$, and $\Omega_{\rm K}=0.$). We show the MCMC samples assuming a flat Universe in Figure \ref{fig:constraints} from the complete 100 Monte Carlo tests and the median, 16\% and 84\% in Table \ref{tab:results}. The posteriors from the other two runs can be found in Appendix \ref{sec:finalResults} (Figure \ref{fig:fixedCosmo} and Figure \ref{fig:allFree}).

Finally we fold in the statistical variance of our PCA estimator from Figure \ref{fig:pcaAnalysis} of $2\%$, adding in quadrature and find, assuming a flat Universe, that  $H_0=71^{+2}_{-3}$km/s/Mpc, $\zl = 0.36_{-0.09}^{+0.2} $, $\alpha=-1.8_{-0.1}^{+0.1}$,   $\mfive=11.1_{-0.1}^{+0.1} $, $\Omega_{\rm M} = 0.3_{-0.04}^{+0.04} $ and $\Omega_{\rm \Lambda}=0.7_{-0.04}^{+0.04}$. 

Following this we estimate the variance due to the choice of noise parameter in the GPR. Although we chose the noise parameter with the highest log-likelihood, this choice could be seen as relatively arbitrary. We therefore measure $H_0$ for a variety of different noise-levels of the GPR. Figure \ref{fig:noiseDep} shows each estimate relative to the fiducial value of $\alpha_{\rm N}=4\times10^{-3}$. We find the estimate is stable within $\sim0.1\%$.

\fig
 \includegraphics[width=0.5\textwidth]{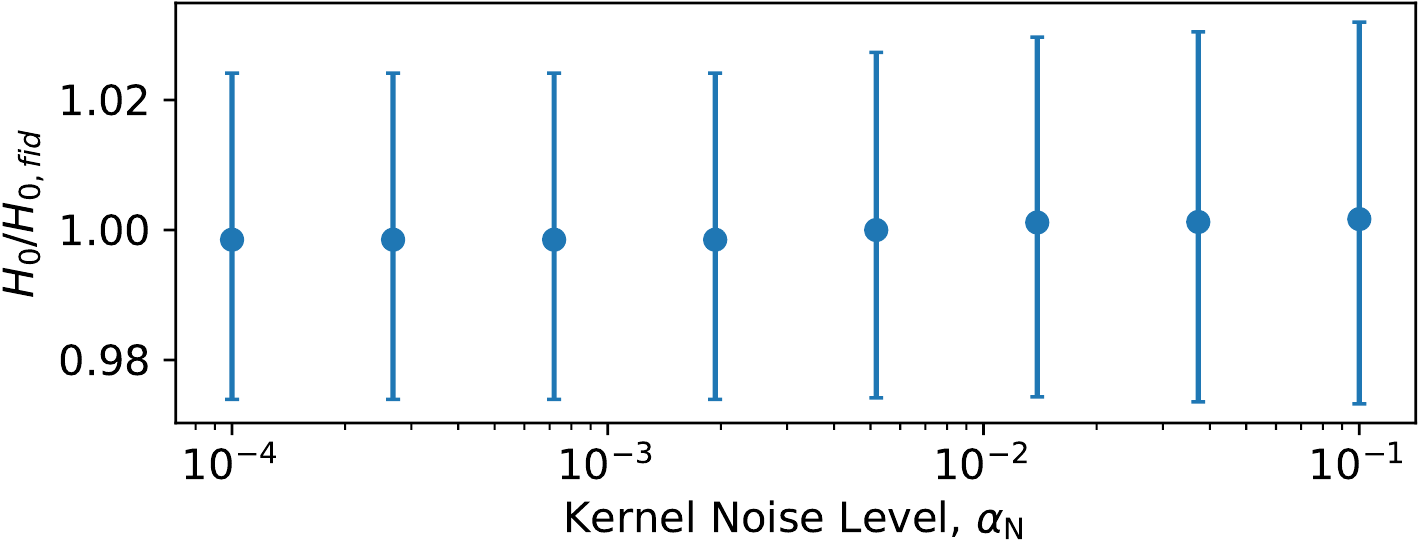}
\caption{\label{fig:noiseDep} {\it Dependence of the estimate of $H_0$ on the GPR meta-parameter $\alpha_{\rm N}$.} We constrain the data using a variety of different noise levels in the Gaussian Processor and find that the value varies within $\sim0.1\%$ around the fiducial estimate ($\alpha_{\rm N}=4\times10^{-3}$). }
\efig


\subsubsection{Appropriateness of the simulations}
In this study we have used the simulated halos to estimate the constraints from data. A key systematic will be the appropriateness of these simulations with respect to the observed lensed systems. We intentionally did not input any lens redshift or mass within $5$kpc in to this method for two reasons. The first was that in the case of large scale surveys this information may not be available, so as such we wanted to test this method as it would on large scale data. Secondly, it would provide an important consistency check. 

This consistency test entails two questions: {\bf 1.} Is the estimated redshift consistent with the true lens redshift? {\bf 2.} Is the estimated mass within $5$ kpc consistent with the same observed mass? We make it clear here that these are consistency checks and not validity checks. Figure \ref{fig:consistency} shows the results of this consistency test. The top panel shows the true lens redshift distribution in red and the estimated distribution from the data. We see that the sampled distribution is bi-modal, with a preference for lower redshifts, however, with also some excess probability at higher redshifts. We hypothesis that the sampler is trying to fit to the high and low redshift regions of the data, however in both cases finds a consistent $H_0$ value.

The bottom panel of Figure \ref{fig:consistency} shows the estimated mass within $5$kpc and the ``true'' mass. In order to estimate the ``true'' observed mass we have make some assumptions about the systems. We first assume that all systems have an image separation of either $r_{\rm sep} = 1.5$ arc-seconds or $r_{\rm sep} = 1.0$ arc-seconds. If we examine the Appendix A in \cite{allTimeDelays}, we see that this is typically the observed separation of the  images. From this we can calculate the mass within the Einstein radius. We note that this is sometimes larger or smaller than $5$kpc, so assuming the $M\propto r$, we account for this difference. We show the estimated mass within $5$ kpc in green and the ``true'' mass assuming a typical image separation of $r_{\rm sep}=1(1.5)$'' in red (green). We see that the estimated masses from these simulations bookend the estimated mass. Moreover, we see that the estimated mass has a broad peak and therefore, for this sample size, we are insensitive to any discrepancies on this scale. As such we state the bias induced by the differences in mass between the observed and simulated sample is small.

\fig

 \includegraphics[width=0.5\textwidth]{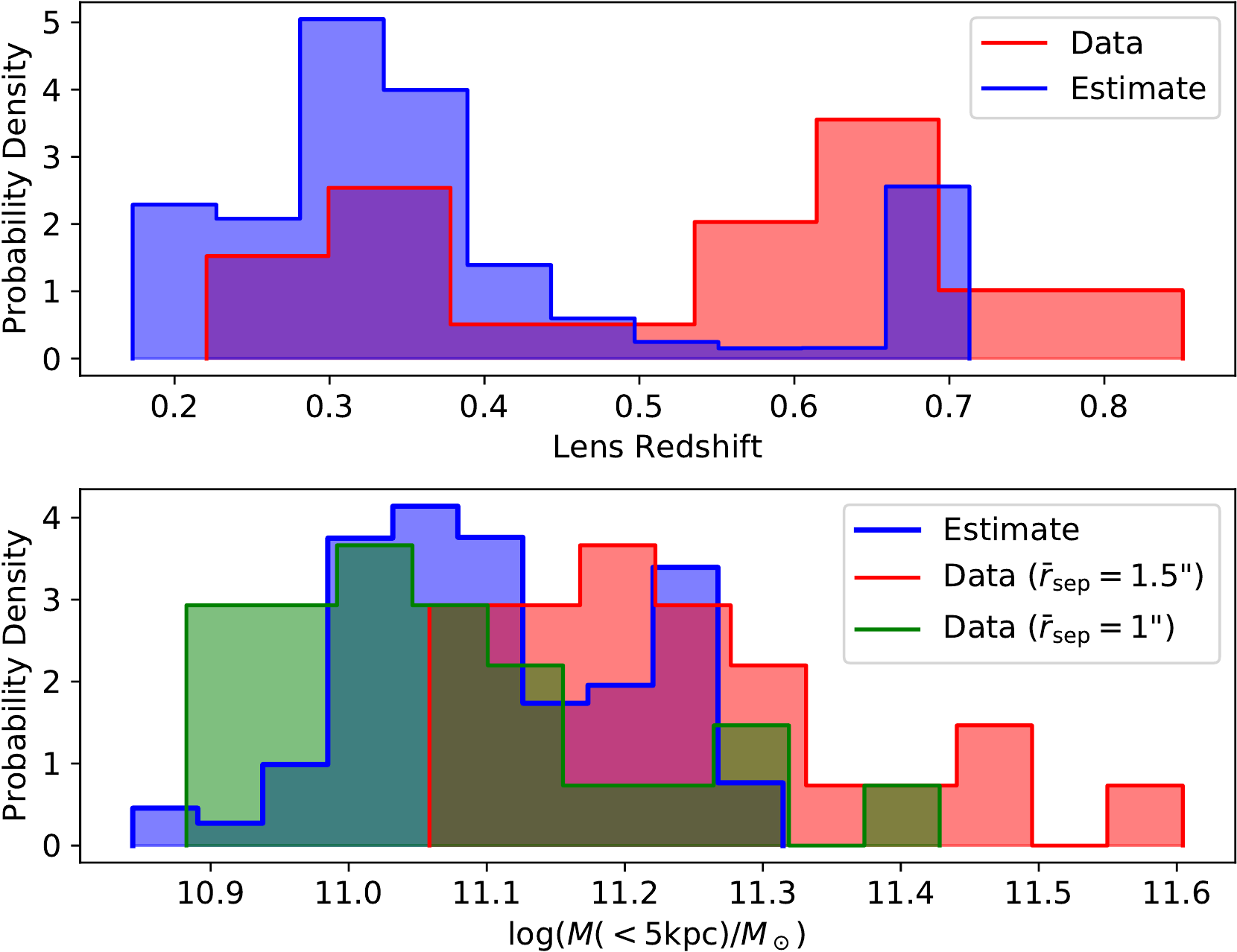}
\caption{\label{fig:consistency} {\it Consistency checks from our fit to the observed data.} The top panel shows the true lens redshift (red) and the posterior from the sampling (blue). We see that the two distributions are consistent with one another and in-fact there is evidence for bi-modality in the posterior, which may reflect the true observed bi-modality. The bottom panel shows the posterior of the mass (blue) and the ``true'' mass assuming a typical image separation of $r_{\rm sep}=1.5 (1.0)$'' in red (green). For these typical image separations the ``true'' mass  is consistent with our mass estimate.}
\efig

\subsection{Statistical reach of the Vera Rubin Observatory}\label{sec:forecasts}

Now we have attempted to measure $H_0$ from current data we explore the power of future data, specifically Vera Rubin Observatory (VRO). In our previous tests we assumed that the observing strategy of VRO is infinite, whereas in truth VRO will have a implicit minimum possible observed time delay due to the return schedule (i.e. how often  a single patch of sky is returned to). As such we now implement a minimum time delay on the CDF of $\Delta t_{\rm min}=10~$days \citep{observingStrategyLSST}.  

Using the same mock samples as we did to the validate the code, we now determine the expected statistical reach of VRO. To do this we estimate the error on each of the six parameters (omitting $\Omega_{\rm K}$ as we are insensitive to this) for a range of lens sample sizes assuming a VRO minimum time delay  (i.e. $\Delta t_{\rm min}=10$). For each case we mock a sample CDF and then fit to the data, iterating 100 times per sample size. Figure \ref{fig:minimumTime} shows the results for each of the six parameters with the error-bar showing the $1-\sigma$ uncertainty in the estimated sensitivity. In the top panel we show the constraints on $H0$, in the dashed cyan line we show the current sensitivity, the red line shows the current sensitivity limit of the GPR model to predict the CDFs. In each panel we show the expected sensitivity from an VRO survey with an optimistic $\sim3000$ lenses and conservative $\sim400$ lenses. We find that up to this conservative limit, our model is sufficient and constraints on $H_0<3\%$ are possible. Similarly marked improvements in the sensitivity in $\Omega_{\rm M}$ and $\Omega_{\rm \Lambda}$ are seen, however these will never be as competitive as other probes. We also find that we maybe able to make significant improvements in our understanding on the global density profile index, with predicted constraints of $<4\%$. 

It is clear from these results that although this model we have presented here is sufficient for the current dataset, it needs improving if we are to exploit the full statistical power of VRO.

\fig
\includegraphics[width=0.49\textwidth]{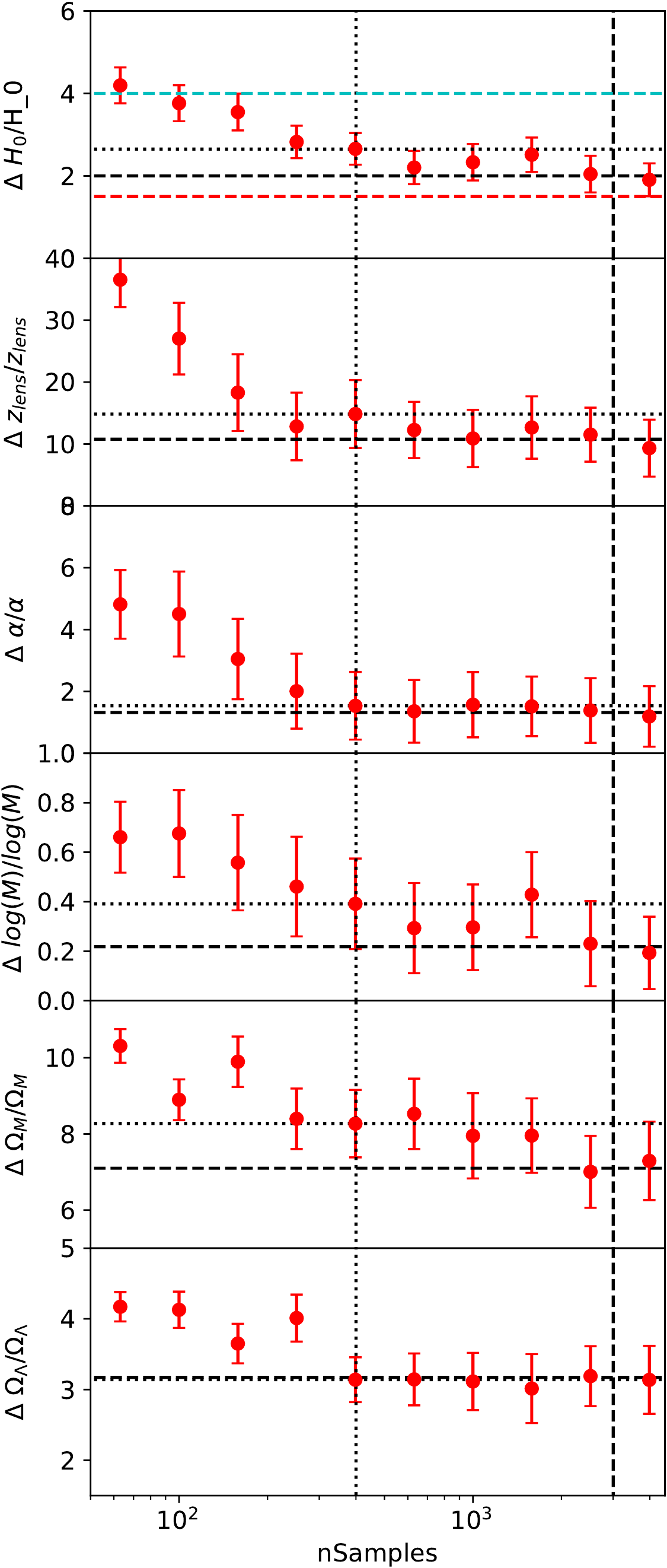}
\caption{\label{fig:minimumTime}  Predicted constraints (percentage) on the six parameters using this method as a function of sample size with a minimum observable time delay of $\Delta t_{\rm min}=10$days.  In the top panel we show the current systematic floor of $\sim2\%$ given by the red line. We also show the current sensitivity from this study in the dashed cyan line. In each panel we should the expected constraints that the Vera Rubin Observatory will gain with an optimistic $\sim3000$ lenses (dashed black) and a conservative 400 lenses (dotted black).  We see that there are immediate gains up to 400, however the accuracy of this model prevents us from predicting constraints lower than this. }
\efig

\begin{table*}
\begin{center}
\caption{\label{tab:results} The three cosmological parameter fitting MCMC runs. We run three different fits, a fixed where we assume $\Omega_{\rm M}=0.3$ and $\Omega_\Lambda=0.7$ and $\Omega_{\rm K}=0$, our fiducial run where we assume a flat Universe, and a final run where we assume nothing about cosmology. }
\begin{tabular}{cccccccc}
\hline
Model&$H_0/($km s$^{-1}$Mpc$^{-1}$)&$z_{lens}$&$\alpha$&log$(M(<5$kpc$)/M_\odot)$&$\Omega_M$&$\Omega_\Lambda$&$\Omega_K$\\
\hline
Fixed& $ 70_{-2}^{+2} $ & $ 0.36_{-0.04}^{+0.09} $ & $ -1.9_{-0.08}^{+0.08} $ & $ 11_{-0.2}^{+0.04} $ \\ 
$\Lambda$CDM& $ 71_{-3}^{+2} $ & $ 0.36_{-0.09}^{+0.2} $ & $ -1.8_{-0.1}^{+0.08} $ & $ 11_{-0.08}^{+0.1} $ & $ 0.3_{-0.04}^{+0.04} $ & $ 0.7_{-0.04}^{+0.04} $ \\ 
$\Lambda$CDMk& $ 71_{-3}^{+3} $ & $ 0.36_{-0.07}^{+0.2} $ & $ -1.8_{-0.1}^{+0.08} $ & $ 11_{-0.08}^{+0.1} $ & $ 0.3_{-0.05}^{+0.05} $ & $ 0.7_{-0.05}^{+0.05} $ & $ 0.00029_{-0.02}^{+0.02} $ \\ 
\hline
\end{tabular}
\end{center}
\end{table*}

\subsection{Discussion}

Here we have presented a $4\%$ statistical error estimate on $H_0$. We have shown that the distribution in lens redshifts and masses are consistent with simulated lenses. Moreover, the method is not too sensitive to the mass parameter and hence although important, will not have a significant bias on the Hubble parameter estimate. However, we do note that this study was based on the simulations of four elliptical galaxies that do not represent the entire population. Moreover, there are known inconsistencies between simulated and observed galaxies \citep{simulatingGalaxies,simulatingGalaxiesA,simulatingGalaxiesB}, particularly in the core where time delays are sensitive. As such, if this method is to be competitive going forward in to the era of sub $2\%$ estimates on $H_0$, then we will require not only more accurate simulations, but an understanding of the time delay distributions over a larger population than just four volumes. Moreover, we showed that the algorithm used here had a $2\%$ systematic error when compared to analytical profiles. If we are to $H_0$ in this fashion in the era of Vera Rubin Observatory then this will need to be improved. Finally in this study we suggest that the galaxies have a shallower profile that that of a Singular Isothermal Sphere. Previous studies  of massive ellipticals from gravitational lensing (e.g. \citep{SLACS_SL}) suggest that galaxies follow a profile closer to that of an singular isothermal sphere ($\alpha=2.$). However this could be due to the fact the GPR was trained on galaxies with profiles $\alpha<2$. Therefore going forward, the finding here would have to corroborated with independent studies as any bias could impact the estimated value of $H_0$. If so it would have implications for the training sample used when estimating the CDFs. Having said that, this method is able to recover the lens redshift of the observed sample, providing some evidence for the potential competitive nature of estimating $H_0$ from a distribution of time delays.

\section{Conclusions}\label{sec:conc}
Strong gravitational time delays are an independent and competitive way to constrain the Hubble Constant, $H_0$.
Studies of individual time delays have led to tight constraints on $H_0$, however in the advent of large scale surveys, this will be difficult to scale. As such we build on previous statistical studies of lensed quasars and propose a new, complementary forward-modelling method to constrain $H_0$ using the Cumulative Density Function (CDF) of observed time delays that does not rely on the details of individual lenses.

We develop a ray-tracing algorithm that is able to take a projected surface density map, for a given lens-source configuration and estimate the CDF. We validate this code on analytical forms of the CDF using Singular Isothermal Spheres. We find that the code can return the expected CDF to within a theoretical systematic error of $\sigma_{\rm sys}<2\%$.

Using full {\it n}-body simulations we use this framework to estimate the CDF of strong gravitational time delays in doubly imaged quasars. In doing so for the first time we do not assume analytical profiles, taking in to account dark halos and substructure and the direct impact of baryonic feedback on time delays.
We include the impact of line-of-sight structures both within the immediate environment of the lens (down to the mass resolution of the simulation) and uncorrelated mass along the line-of-sight, micro-lensing and magnification bias.

We find that the CDFs exhibit large amounts of halo to halo variance, caused by varying number of substructures at different redshifts, projected density profiles and total mass with $5$ kpc. Indeed we find that the the relation between the power law index of the probability density function (PDF) and the density profile is consistent with analytical expectations and that the most probable time delay (i.e. the peak of the PDF) is proportional to the square of the total mass within $5$kpc, again consistent with analytical expectations. However, there is a large amounts of scatter around these relations, showing how individual halos can induce different features in the CDFs.

We construct a model of the CDF for a fiducial cosmology by reducing their dimensionality using a Principal Component Analysis and then interpolating between these using a Gaussian Process Regressor. We then calculate the cosmological shift from different parameter sets and fit a linear model.  This way we are able to predict a CDF for a given  lens redshift, $\zl$, power law index, $\alpha$ and a full set of cosmological parameters including $H_0$, $\Omega_{\rm M}$, $\Omega_{\rm \Lambda}$ and $\Omega_{\rm K}$. We carry out a self-consistency test estimating six parameters (omitting $\Omega_{\rm K}$ since we are insensitive to this). We find the model returns the expected value however there is degeneracy between the density profile parameter $\alpha$ and the mass within $5$ kpc. We also find that all cosmological parameters are mildly degenerate with one another. We then estimate the true $H_0$ from a sample of 27 doubly imaged time delays reported in \cite{allTimeDelays}. Assuming a flat Universe, we measure    $H_0=71^{+2}_{-3}$km/s/Mpc, $\zl = 0.36_{-0.09}^{+0.2} $, $\alpha=-1.8_{-0.1}^{+0.1}$,   $\mfive=11.1_{-0.1}^{+0.1} $, $\Omega_{\rm M} = 0.3_{-0.04}^{+0.04} $ and $\Omega_{\rm \Lambda}=0.7_{-0.04}^{+0.04}$. This amounts to a $4\%$ estimate of the Hubble parameter. 

We discuss the appropriateness of the simulations, and find that within the sensitivity of current data we are not systematically biased. We return a consistent lens redshift distribution and the estimated masses are both within the expected true mass, and that the model is not particularly sensitive to this parameter. However, we note that there is debate on the reliability of such simulations at reproducing the properties of galaxies, which may impact these results \citep{simulatingGalaxies,simulatingGalaxiesA,simulatingGalaxiesB}.

Finally, we estimate the statistical reach of future data, specifically Vera Rubin Observatory (VRO). We find that large improvements can be made in the first few hundred lenses, with predicted constraints of $\Delta H_0/H0<3\%$. However, due to the sensitivity limit of the model ($2\%$), beyond $1000$ lenses, the model will need to be improved before advances can be made. Should VRO return the optimistic $3000$ lenses, then an improved model, trained on many more simulations, could make a competitive measurement of the Hubble Constant. 

\bsp

\section*{Acknowledgments}
We would like to thank Richard Massey, Frederic Courbin and Alessandro  Sonnenfeld for valuable conversations and insight. We would also like to thank Dr. Benjamin Oppenheimer for contributing his simulations of Cold Dark Matter, without which we would not have been able to complete this project. DH is supported by the D- ITP consortium, a program of the Netherlands Organization for Scientific Research (NWO) that is funded by the Dutch Ministry of Education, Culture and Science (OCW)

\section*{Data Availability}
Data is private, however access is available upon request.
\label{lastpage}
\bibliographystyle{mn2e}
\bibliography{bibliography}
\appendix
\section{Time delay sample}\label{sec:timeDelays}
\begin{table*}
\caption{Over of all time delay measurements from doubly image quasars taken from the literature.
{\it Col 1:} Quasar ID, {\it Col 2:} Measurement reference from literature, {\it Col 3:} Redshift of the lens, {\it Col 4:} Redshift of the source, {\it Col 5:} Time delay with associated error.}
\label{tab:timeDelays}
\begin{center}
\begin{tabular}{l c c c c }
\hline
\hline
Quasar ID  & Ref & $z_{\rm lens}$ & $z_{\rm source}$ & $\Delta t$   \\
                   &         &          &            & (days)                    ) \\
   \hline                
HE 0047-1756 & \cite{allTimeDelays} &  0.407 & 1.678 &$10.4\pm3.5$ \\ 
\hline
Q 0142-100 & \cite{allTimeDelays} & 0.491& 2.719  &$ 97_{-15.5}^{+16.1}$\\ 
\hline
Q J0158-4325 & \cite{allTimeDelays} &0.317& 1.29   &$22.7\pm3.6$\\ 
\hline
SDSS J0246-0825 & \cite{allTimeDelays} &0.723&1.689  &$0.8_{-5.2}^{+5.0}$\\ 
\hline
HS 0818+1227 & \cite{allTimeDelays} &0.39& 3.115  &$153.8_{-14.6}^{+13.2}$\\ 
\hline
SDSS J0832+0404 & \cite{allTimeDelays} &0.659& 1.115  &$125.3_{-23.4}^{+12.8}$\\ 
\hline
SDSS J1226-0006 & \cite{allTimeDelays} &0.517& 1.123  &$33.7\pm2.7$\\ 
\hline
Q 1355-2257 & \cite{allTimeDelays} &0.701&1.370 & $81.5_{-12}^{+10.8}$\\ 
\hline
SDSS J1455+1447 & \cite{allTimeDelays} & NA & 1.424  & $47.2_{-7.8}^{+7.5}$\\ 
\hline
SDSS J1515+1511 & \cite{allTimeDelays} &0.742& 2.054  & $210.2_{-5.7}^{+5.5}$\\ 
\hline
SDSS J1620+1203 & \cite{allTimeDelays} & 0.398 & 1.158 & $171.5\pm{8.7}$ \\ 
\hline
HE 2149-2745& \cite{allTimeDelays} & 0.603  & 2.033 &$39_{-16.7}^{+14.9}$\\ 
 \hline
Q0142$-$1002 & \cite{Koptelova2012} & 0.491 & 2.719 & $89 \pm 11$ \\
\hline
JVAS B0218+357& \cite{Cohen2000} & 0.685 & 0.944 & $10.1_{-1.6}^{+1.5}$ \\
\hline
SBS 0909+532& \cite{Goicoechea2008} & 0.830 & 1.377 &  $60_{-4}^{+2}$ \\
\hline
FBQ 0951+2635 &\cite{Jakobsson2005} &0.260 & 1.246 & 16 $\pm$ 2  \\
\hline
SDSS J1001+5027 &\cite{Kumar2013} &  0.415& 1.838  & 119.3 $\pm$ 3.3 \\
\hline
HE 1104$-$1805 &\cite{Poindexter2007} & 0.729  &  2.319 & $152.2^{+2.8}_{-3.0}$ \\
\hline
SDSS J1206+4332& \cite{Eulaers2013} & 0.748 & 1.789 & 111.3 $\pm$ 3 \\
\hline
SBS 1520+530& \cite{Burud2002b} &  0.717 &  1.855 & 130 $\pm$ 3  \\
\hline
B1600+434 &\cite{Burud2002a} & 0.414 & 1.589 & 51 $\pm$ 4 \\
\hline
SDSS J1650 + 4251 &\cite{Vuissoz2007} &  0.577 & 1.547 & 49.5 $\pm$ 1.9 \\
\hline
PKS 1830$-$211& \cite{Lovell1998} &  0.89 & 2.507 & 26$^{+4}_{-5}$  \\
\hline
HE 2149$-$2745 &\cite{Burud2002a} & 0.603& 2.033 & 103 $\pm$ 12 \\
\hline
HS 2209+1914& \cite{Eulaers2013} &0.68 & 1.07 & 20.0 $\pm$ 5  \\
\hline
SDSS J1339 + 1310 &\cite{Goicoechea2016} & 0.609 & 2.243 &  $47^{+5.0}_{-6.0}$ \\
\hline
SDSS J1442 + 4055 & \cite{Shalyapin2019} & 0.284 & 2.0 & $25\pm1.5$ \\ 
\hline
\end{tabular}
\end{center}
\end{table*}

\section{Other cosmological fits}\label{sec:finalResults}
Here we show the posteriors from the other two MCMC runs assuming a fixed cosmology (i.e. $\Omega_{\rm M}=0.3$, $\Omega_{\rm \Lambda}=0.7$, $\Omega_{\rm K}=0.$) and a completely free cosmology.
\figs
 \includegraphics[width=\textwidth]{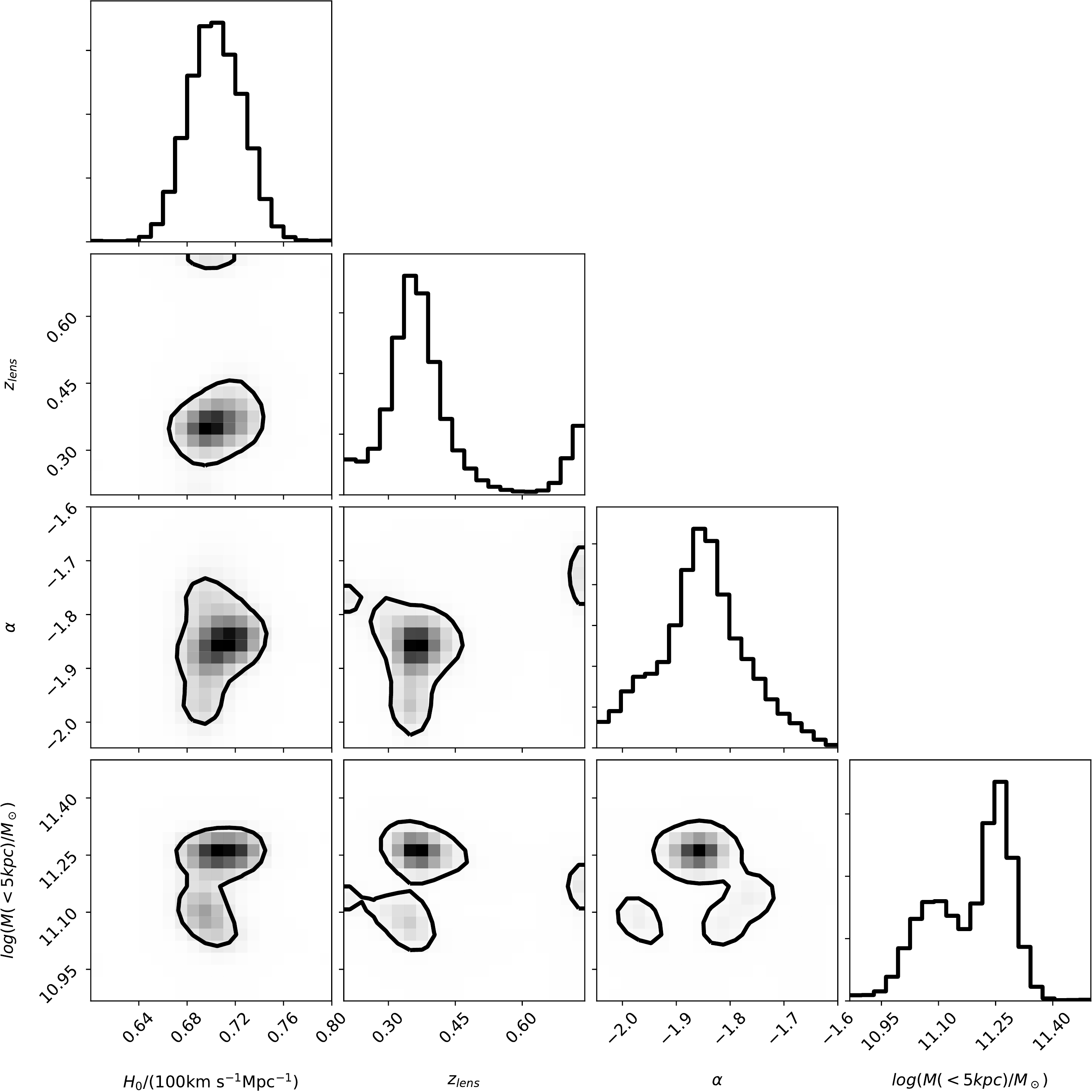}
\caption{\label{fig:fixedCosmo}  Posterior from a fixed cosmological model (i.e. $\Omega_{\rm M}=0.3$, $\Omega_{\rm \Lambda}=0.7$, $\Omega_{\rm K}=0.$)}
\efigs

\figs
 \includegraphics[width=\textwidth]{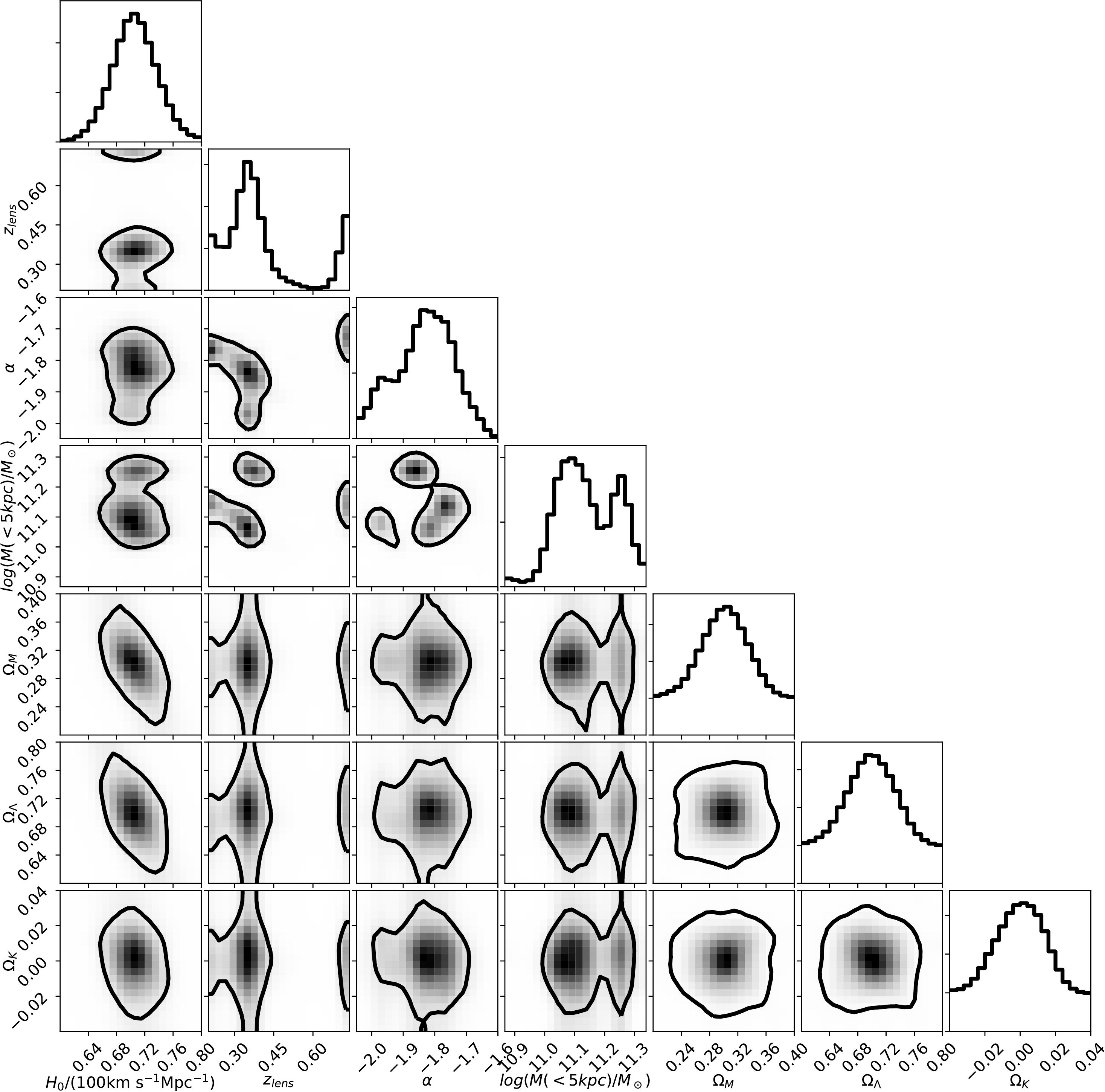}
\caption{\label{fig:allFree} Posteriors from an open cosmology.}
\efigs

\end{document}